\documentclass[openacc]{rstransa}%%%%where rsproca is the template name

\usepackage{lmodern}
\usepackage{url,hyperref,lineno,microtype,subcaption}
\usepackage{graphicx}% Include figure files
\graphicspath{{figures/}} %Setting the graphicspath
\usepackage{amsmath}
\usepackage{amssymb}
\usepackage{braket}
\usepackage{physics} %abs
\usepackage{mathtools} %coloneq
\usepackage{cancel}
\titlehead{Research}

\begin{document}

\jname{rsta}
\Journal{Phil. Trans. R. Soc. A}

%%%% Article title to be placed here
\title{Second-order optimisation strategies for neural network quantum states}

\author{%%%% Author details
M.~Drissi$^{1}$,
J.~W.~T.~Keeble$^{2}$,
J.~Rozalén Sarmiento$^{3,4}$ and 
A.~Rios$^{2,3,4}$}

%%%%%%%%% Insert author address here
\address{
$^{1}$TRIUMF, Vancouver, V6T 2A3, British Columbia, Canada
\\
$^{2}$ Department of Physics, University of Surrey, Guildford GU2 7XH, United Kingdom
\\
$^{3}$ Departament de F\'isica Qu\`antica i Astrof\'isica, 
 Universitat de Barcelona (UB), 
 c. Mart\'i i Franqu\`es 1, E08028 Barcelona, Spain
\\
$^{4}$ Institut de Ci\`encies del Cosmos (ICCUB),
Universitat de Barcelona (UB), 
Barcelona, Spain
\\
\includegraphics[keepaspectratio,width=1em]{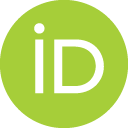}
MD, \href{https://orcid.org/0000-0001-9472-6280}{0000-0001-9472-6280};
JWK, \href{https://orcid.org/0000-0002-6248-929X}{0000-0002-6248-929X};
JRS, \href{https://orcid.org/0000-0002-0660-1216}{0000-0002-0660-1216};
AR, \href{https://orcid.org/0000-0002-8759-3202}{0000-0002-8759-3202}
}

%%%% Subject entries to be placed here %%%%
\subject{quantum physics, nuclear physics, artificial
intelligence, game theory}

%%%% Keyword entries to be placed here %%%%
\keywords{neural network quantum states, Variational Monte Carlo, game theory, decision geometry, optimisation}

%%%% Insert corresponding author and its email address}
\corres{M.~Drissi\\
\email{mdrissi@triumf.ca}}

\newcommand*{\transpose}{{\mkern-1.5mu\mathsf{T}}} % for a nice transpose symbol

\begin{abstract}
The Variational Monte Carlo method has recently seen important advances through the use of neural network quantum states.
While more and more sophisticated ansätze have been designed to tackle a wide variety of quantum many-body problems,
modest progress has been made on the associated optimisation algorithms.
In this work, we revisit the Kronecker-Factored Approximate Curvature, an optimiser
that has been used extensively in a variety of simulations.
We suggest improvements on the scaling and the direction of this optimiser, and find that they
substantially increase its performance at a negligible additional cost.
We also reformulate the Variational Monte Carlo approach
in a game theory framework, to propose a novel optimiser based on decision geometry.
We find that, on a practical test case for continuous systems, this new optimiser consistently outperforms any of the
KFAC improvements in terms of stability, accuracy and speed of convergence.
Beyond Variational Monte Carlo, the versatility of this approach suggests
that decision geometry could provide a solid foundation for accelerating 
a broad class of machine learning algorithms.
\end{abstract}
\maketitle

\begin{fmtext}
\end{fmtext}
\section{Introduction}

Neural network quantum states (NQS) have been recently employed to represent the ground-state wave functions of several quantum many-body systems~\cite{carleo2017solving,hermann2023,pfau2020ab,hermann2020deep,spencer2020better,von2022self,keeble2023machine,pescia2023message,rigo2023solving,kim2023neuralnetwork,lou2023neural}.
These state-of-the-art machine learning (ML) tools allow for numerically efficient solutions of the Schr\"odinger equation, while keeping the 
relevant information of the wave function in a finite, relatively small number of parameters.
The primary focus of NQS has been to leverage advancements in ML
to represent the widest class of many-body wave functions.
The first NQS was developed in Ref.~\cite{carleo2017solving}, where a Restricted Boltzmann Machine
was used to model spin systems.
More recently, the field of quantum chemistry has seen a surge of ans\"atze representing
the wave function of many-electron systems~\cite{hermann2023,pfau2020ab,hermann2020deep,spencer2020better,von2022self}.
Message-passing architectures 
have also been proposed to tackle strongly correlated fermionic systems~\cite{pescia2023message,lou2023neural,kim2023neuralnetwork}.
A key ingredient of many of these NQS
is the use of permutation-equivariant layers. These 
provide an efficient representation of the required antisymmetry of fermionic many-body wave functions.

To tackle the quantum many-body problem, NQS are usually formulated in the framework of 
Variational Monte Carlo (VMC)~\cite{becca_sorella_2017,martin_reining_ceperley_2016}.
Like any variational approach, VMC calculates the ground-state of a many-body system by minimising
the expectation value of the energy of an associated Hamiltonian over a class of trial wave functions. 
Provided the class is large enough, this approach should yield a good approximation of
the ground state, even in the case of strongly interacting particles.
Monte Carlo techniques are employed to estimate the many-body, and generally high-dimensional, integral 
for the ground-state energy. 
In practice, with NQS, one reformulates the physics energy minimisation process 
in terms of a ML reinforcement learning problem. 
The loss function is the total energy of the system, and the wave function 
is learned by minimising it.

In nuclear physics, specifically, NQS have shown promising results. Starting from a simple solution to the deuteron~\cite{Keeble2020}, the field has quickly evolved towards fully-fledged nuclear structure calculations~\cite{Adams2021,gnech2022,gnech2023}. Further
advances in architectures have opened the window to applications in periodic systems~\cite{Pescia2022}  
like neutron matter~\cite{fore2023}, and in models based on the  
occupation number formalism~\cite{rigo2023solving}. 
The NQS reformulation may be able to 
surpass previous efforts in VMC simulations of nuclear systems~\cite{Carlson2015,Lomnitz1981}.
While simulations have not yet reached
the level of sophistication of other 
\emph{ab initio} methods~\cite{Hagen2014,Hergert2016,Soma2020}, NQS have 
the potential to become a new and competitive 
first-principles technique.

Clearly, the optimisation algorithm in this minimisation process plays a critical role to achieve
results efficiently. 
In the most challenging cases, typically involving many particles, 
the wall-time per epoch may increase substantially. 
With these cases in mind, one may want to design optimisation strategies
that reduce the total number of epochs while keeping a reasonable wall-time per epoch.
Typically, optimisation strategies are based on gradient descent methods,
which exploit the first-order derivatives of the loss function 
with respect to the variational parameters.
In a ML setting, derivatives are efficiently estimated using automatic differentiation (AD) techniques~\cite{baydin2018automatic}.
A possible approach to accelerate the optimisation consists in approximating second-order 
derivatives of the loss function to incorporate information on the local curvature
in the update of the parameters.
This typically leads to a trade-off, where the total number of epochs to converge accurately
is reduced at the price of increasing the wall-time per epoch.

The Adam~\cite{kingma2014adam} and Stochastic Reconfiguration (SR)~\cite{sorella1998green, sorella2001SRforVMC} optimisers have been extensively used in the NQS community. While Adam provides a stable and robust optimisation strategy, its convergence to accurate results is generally fairly slow. As an example, Refs.~\cite{pfau2020ab,keeble2023machine} indicate that Adam requires between $10^4$ to $10^5$ epochs to reach an accuracy of $5\%$ in the correlation energy.
In contrast, SR has been shown to work remarkably well and has become a key factor in VMC calculations \cite{carleo2017solving, rigo2023solving, pescia2023message, Adams2021, gnech2022, Park2020nqs}.

The Kronecker-Factored Approximate Curvature (KFAC)~\cite{martens2015optimizing} optimiser has also been shown to perform notably well~\cite{martens2015optimizing,grosse2016kronecker,martens2020new} and has been essential in VMC calculations for some of the most challenging cases~\cite{pfau2020ab,spencer2020better,von2022self,wilson2021simulations}.
However, the KFAC optimiser was originally designed for supervised learning 
problems~\cite{martens2015optimizing} and has mostly
been tested with standard feed-forward neural networks (NN). 
Despite several extensions, for example to include
convolutional layers~\cite{grosse2016kronecker}, the justification of its applicability to VMC with 
NQS remains unclear. As a matter of fact, to bypass problems encountered with KFAC in VMC, several groups have come up
with a series of \emph{ad hoc} adjustments to rectify \emph{a posteriori} the update on the parameters
obtained from KFAC~\cite{pfau2020ab}. 
These adjustments go against the philosophy originally advocated in Ref.~\cite{martens2015optimizing}, 
which suggested that the learning rate and momentum should be automatically adjusted based on the exact Fisher matrix,
and where the damping was dynamically adjusted using a Levenberg-Marquardt rule.

In this work, we revisit the original KFAC optimiser and review its formulation within a VMC-NQS framework. %calculations with NQS.
We aim to tackle its shortcomings at the very basis, in order to build a new versatile optimiser that may
accelerate VMC calculations for a wide variety of many-body problems.
We introduce in Sec.~\ref{sec:quantumManyBody} the quantum many-body problem that we use as a test bed: strongly-interacting polarised fermions in a one-dimensional harmonic trap.
The NQS ansatz used in VMC and the general KFAC algorithm are also briefly discussed.
Then, in Sec.~\ref{sec:adaptingKFAC}, we single out several important issues that we encountered with 
the original implementation of KFAC.
We show how these relate to fundamental aspects of VMC and NQS, namely the use of the energy as a loss function
and the use of permutation equivariant layers. From this analysis, we build a new, improved
KFAC-like optimiser that addresses directly these issues and shows a dramatic improvement in terms of
stability and accuracy.
Finally, in Sec.~\ref{sec:decisionalGradient}, we revisit the underlying
information geometry~\cite{amari1983foundation,amari2016information} on which KFAC
is built~\cite{martens2015optimizing}.
We reformulate the VMC problem in a game theory setting, which motivates an alternative decision geometry~\cite{dawid2005geometry,dawid2007geometry,gneiting2007strictly}.
A novel optimiser based on this decision geometry is then introduced, which outperforms any of the previously tested optimisers.

\section{Quantum many-body model}
\label{sec:quantumManyBody}

\subsection{System and Hamiltonian}

In the following, we provide a general discussion on optimisation strategies for quantum many-body systems.
For concreteness and pedagogical purposes, several aspects of the discussion are illustrated with a simple many-body system~\cite{keeble2023machine}. 
The toy model we consider is a system of $A$ fermions of mass $m$,
trapped in a one-dimensional harmonic well~\cite{keeble2023machine}.
For simplicity, we assume the system to be fully polarised and thus neglect spin effects.
We employ Harmonic Oscillator (HO) units and choose a Gaussian-shaped two-body interaction with range
$\sigma_0$ and strength $V_0$, inspired by nuclear-physics applications~\cite{keeble2023machine}.
The Hamiltonian of the system thus reads
\begin{align}
H =& - \frac{1}{2} \sum_{i=1}^A  \nabla_i^2 + \frac{1}{2} \sum_{i=1}^A  x_i^2  + 
\frac{V_0}{\sqrt{ 2 \pi} \sigma_0} \sum_{i<j}  e^{ - \frac{ (x_i-x_j )^2}{2 \sigma_0^2} } \ .
\label{eq:hamiltonian}
\end{align}
Throughout this work, we use this system as a toy model to benchmark the performance of different energy
optimisation schemes.
Specifically, we change the number of particles from $A=2$ to $6$ and the interaction strength, $V_0$,
from $-20$ to $20$. We keep a fixed interaction range, $\sigma_0=0.5$, which allows for an exploration
of the non-perturbative regime~\cite{keeble2023machine}.

This system has the advantage to be simple while displaying a rich phenomenology.
Its simplicity allows us to explore many different optimisation scenarios while, its phenomenological diversity
supports the fact that our conclusions can not be too specific to the many-body problem we consider.
In particular, when varying $V_0$ between $-20$ and $20$, the many-body system undergoes a transition
from a strongly attractive regime, where spinless fermions behave like a non-interacting bosonic 
system~\cite{Girardeau2003,Granger2004,Valiente2020},
to a strongly repulsive regime, akin to that of a crystal~\cite{wigner1934interaction,Ziani2020}.

\begin{figure*}[t]
\begin{center}
\includegraphics[width=0.8\linewidth, trim= 0 40 0 40]{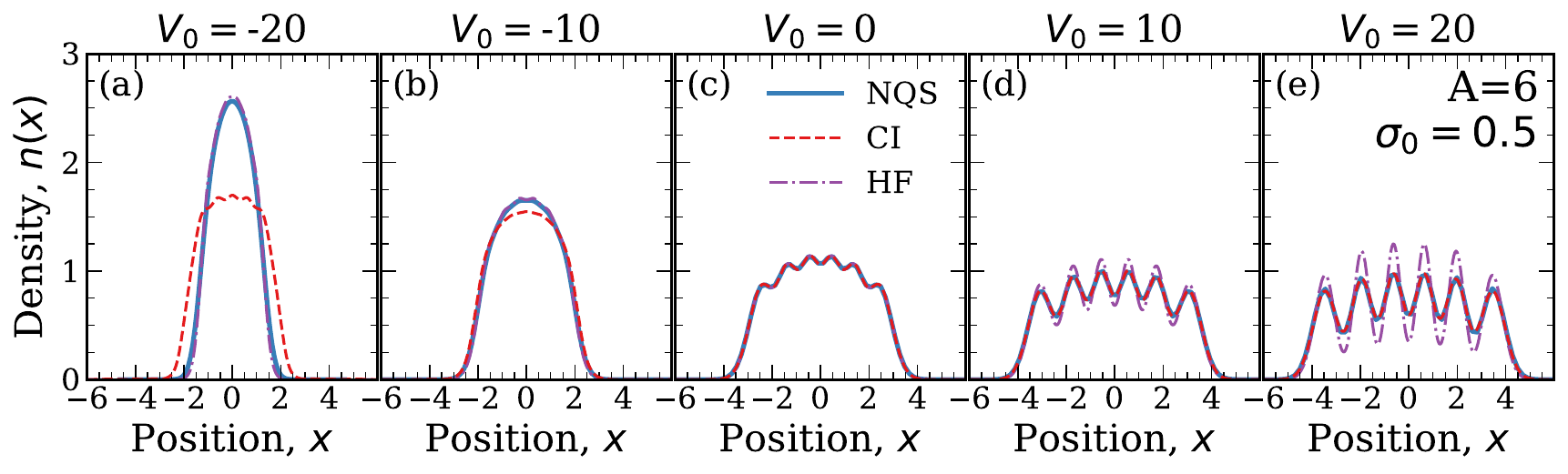}
\end{center}
\caption{The density profile $n(x)$ of $6$-particle system as a function of position $x$.
Panels (a) to (e) show results for values of $V_0$ from $-20$ to $20$ in steps of $10$.
The solid blue line corresponds to the VMC-NQS simulations. Dashed red lines show the 
direct diagonalisation (CI) results, whereas the purple dashed-dotted line displays the 
Hartree-Fock (HF) approximation.}
\label{fig:densityProfileExample}
\end{figure*}

As an illustrative example, we show in Fig.~\ref{fig:densityProfileExample} the one-body density profile,
$n(x)$,
as a function of the spatial position for the $A=6$ system. 
The different panels correspond to different interaction strengths.
We show our NQS prediction with a solid line, but provide also two additional benchmarks. Dashed lines
correspond to a configuration interaction (CI) result. This employs a full direct diagonalisation of the 
Hamiltonian within a given configuration space~\cite{RojoFrancas2020}.
The associated Hartree-Fock (HF) results are shown with dash-dotted lines. 

The central panel (c) shows the non-interacting case, $V_0=0$. It displays an overall Gaussian-like profile
with a series of $6$ superimposed peaks, which are characteristic of fermion systems. 
Panels (a) and (b), to the left of the centre, show the
attractive regime with $V_0=-20$ and $V_0=-10$, respectively. 
In this region, the wiggles disappear and the density profile acquires a dumbbell shape. This relatively
featureless density profile is reminiscent of a bosonic system~\cite{keeble2023machine}. 
In this regime, the CI results disagree with the NQS prediction because of the 
truncation of the model space, which is performed for practical numerical reasons. 
Here, and in the following, we will
not show CI results for $V_0=-20$ for this reason.
% In contrast, the HF results agree relatively well with the NQS prediction in the attractive regime. 

Panels (d) and (e) correspond to the repulsive regime for values of the interaction strength 
$V_0=10$ and $20$. They suggest a very different behaviour of the density profile
in the repulsive regime. The system does not change substantially in average size, but the fermionic oscillations
on top of the smooth behaviour become much more pronounced. This is a sign of localisation, or crystallisation,
in the system.
% We note that here the HF density profile differs substantially from the NQS result,
% although the energy predictions (not shown here) are relatively close to the NQS predictions.
For more details on the physics of this model, we refer to our previous work in Ref.~\cite{keeble2023machine}.

\subsection{The Variational Monte Carlo method}

To compute the ground state of the Hamiltonian in Eq.~\eqref{eq:hamiltonian} we follow a VMC approach~\cite{becca_sorella_2017}
and aim at solving the many-body problem using the Rayleigh-Ritz variational principle,
\begin{equation}
    \frac{\Braket{ \Psi^A | H | \Psi^A} }{\braket{\Psi^A}{\Psi^A}} \geq E^A_{\text{g.s.}} \ ,
    \label{eq:RR_energy}
\end{equation}
where $E^A_{\text{g.s.}}$ denotes the exact $A$-body ground-state energy of $H$. 
In VMC, the wave function $\ket{\Psi^A}$ is not in general the exact ground state, 
but rather a trial wave function $\ket{\Psi_T}$ that depends on a series of parameters. 
Using the Rayleigh-Ritz variational principle, the original eigenvalue problem can thus be recast as
an optimisation problem in the subspace of trial wave functions. 

In practice, three major challenges need to be faced to solve this optimisation problem.
First of all, the Hilbert space of $A$-body states in the optimisation search is in principle 
infinite-dimensional. This difficulty is usually tackled by introducing the much smaller set of trial states $\ket{\Psi_T}$
over which the energy is minimised. In doing so, a bias is introduced on the solution, which is usually
estimated by varying the size of the set of trial states.
A second difficulty comes from the necessity to evaluate the expectation value of the energy over many 
different trial states.
Each estimate consists of an integral of at least dimension $D \times A$, where $D$ denotes the number of space
dimensions. To mitigate the cost of evaluating these high-dimensional integrals, one 
resorts to Monte Carlo estimates, which necessarily 
introduce a statistical noise on the energy. This can in principle be controlled by using a large enough number
of samples in the Monte Carlo estimate. 
Finally, the third challenge is to find the solution of a global non-linear optimisation problem.
Here, the established strategy consists in breaking the problem into a sequence of simpler,
local-quadratic optimisation sub-problems, which are solved iteratively.
Such iterative optimisers have the advantage of versatility and can be used for many
different optimisation problems. However, they often require adjustments to converge
in a reasonable time~\cite{nocedal2006numerical}.

The third challenge is the main focus of this work, and we will address it in section~\ref{sec:adaptingKFAC}.
Before diving into it, we provide some details on our NQS strategy, which will be useful to frame the global
discussion on the optimisation strategy.

\subsection{Neural Quantum States and Monte Carlo sampling}\label{sec:NQSandMC}

Our NQS ansatz is a fully antisymmetric NN, inspired by the  
implementation of FermiNet~\cite{pfau2020ab}. 
The inputs to our network are the $A$ positions of fermions in the system, $\{ x_i , i = 1, \ldots, A \}$,
and the output is the
many-body wave function $\Psi_\theta(x_1,\ldots,x_A)$ in the log domain.
The state depends on a series of network weights and biases,
succinctly summarised by a variable $\theta$ of dimension $N$~\cite{keeble2023machine}.
The network is composed of $4$ core components:
equivariant layers, generalised Slater matrices (GSMs), Gaussian log-envelope functions, 
and a summed signed-log determinant function~\cite{keeble2022neural}.
The equivariant layers have the particularity of sharing a common set of weights and biases among
the $A$ input positions.
This weight sharing is critical in order to enforce total antisymmetry of the wave function
and, as discussed later, will impact the rationale justifying KFAC.
For all calculations presented in this work, the number of equivariant layers $L$ is set to $L=2$; their width,
$H$, is $H=64$; and we only use one GSM. We stress that the orbitals
making a GSM are functions depending on the positions of \emph{all} the particles, thus departing
from a standard Slater matrix. 
These so-called backflow correlations are key in obtaining a faithful representation of the exact state~\cite{luo2019backflow, keeble2023machine}. 
In total, our wave function ansatz depends on about $10^4$ parameters,
which have to be optimised to solve the many-body problem.

The expectation value of the energy, in Eq.~\eqref{eq:RR_energy}, is typically formulated as a statistical average over \emph{local} energies, 
\begin{align}
    E(\theta) & \equiv \frac{\langle \Psi_\theta \vert H \vert \Psi_\theta \rangle}{\langle \Psi_\theta \vert \Psi_\theta \rangle} =
    \mathbb{E}_{X \sim p_\theta}
    \left[ \Psi_\theta(X)^{-1} H \Psi_\theta(X) \right]  \ ,
    \label{eq:stochastic_energy}
\end{align}
where we define $X=(x_1, \ldots, x_A)$ as an $A$-dimensional 
random variable (or walker) distributed according to the Born 
probability of the many-body wave function, $p_\theta=\abs{\Psi_\theta}^{2}$.
For numerical stability, the kinetic energy is computed in the log domain. 
For a general local interaction, the local energy in a continuous model reads
\begin{align}
E_{L,\theta}(X) & \equiv 
     \Psi_\theta(X)^{-1} H \Psi_\theta(X) =  
     -\frac{1}{2} \sum_{i=1}^A \left[ 
     \partial^{2}_{x_i} \ln\abs{\Psi_\theta(X)}+ 
     \left(\partial_{x_i} \ln\abs{\Psi_\theta (X)} \right)^{2} 
    \right] 
     + V(X)
\ . 
    \label{eq:local_energy}
\end{align}
For our specific model, the interaction terms are
\begin{equation}
    V(X) \equiv \frac{1}{2}\sum_{i=1}^A x_i^2  + 
\frac{V_0}{\sqrt{ 2 \pi} \sigma_0} \sum_{i<j}  e^{ - \frac{ (x_i-x_j )^2}{2 \sigma_0^2} } \ .
\label{eq:interaction}
\end{equation}
To sample the statistical averages over the Born probability $p_\theta$,
we use $N_W=4096$ walkers propagated following a Metropolis-Hastings algorithm
(MH)~\cite{metropolis1949monte,hastings1970monte} with on-the-fly adaptation of
the proposal distribution width~\cite{wilson2021simulations}.
In order to neglect any biases due to autocorrelation in our comparison of multiple optimisers,
we perform $400$ MH iterations per epoch, to ensure the thermalisation of the Markov chain.
The optimisation process requires not only the evaluation of the energy expectation value, $E(\theta)$, but also the gradients with respect to all 
the parameters of the network $\nabla_{\theta}E(\theta)$. These are  evaluated as a statistical average according to
\begin{equation}
    \nabla_{\theta}E(\theta)
        =
        2\mathbb{E}_{X \sim p_\theta}\left[
            \left(E_{L,\theta}(X) - E(\theta)\right)
            \nabla_{\theta}\ln\abs{\Psi_\theta(X)}
        \right] \ .
\end{equation}

Finally, to provide a reasonable starting point to the energy minimisation algorithms,
we drive the NQS through a pre-training phase. 
This is a supervised learning problem, where we demand 
that the network reproduce the many-body wave function of the non-interacting system~\cite{keeble2022neural,keeble2023machine}. 
% The goal is to obtain an initial state that is physical and
% somewhat similar, in terms of spatial extent, to the 
% result after the interaction is switched on. 
% This also helps speed up the 
% simulation in the more demanding 
% cases~\cite{keeble2022neural,keeble2023machine}.
Throughout this work, the parameters of the network are randomly initialised before going through $10^3$
pre-training epochs using the Adam optimiser.
 The final pre-trained NQS is the starting point for all the
 optimisation 
algorithms that we consider. For a given particular system 
(number of particles $A$ and interaction strength $V_0$), we always 
start from the same exact NQS and set of samples. This guarantees that 
any differences that arise in the optimisation process are due to
the optimisers themselves.
We refer the reader to the Supplemental Material and to Refs.~\cite{keeble2022neural,keeble2023machine} for more details on 
the Monte Carlo sampling and the pre-training used in our calculations.

% \section{KFAC within a Variational Monte Carlo framework}
% \label{sec:adaptingKFAC}
\section{Optimisation within a Variational Monte Carlo framework}
\label{sec:adaptingKFAC}

% In this section, we present a series of optimisation strategies that have been employed in past NQS implementations of VMC.
In this section, we present a series of optimisation strategies that we test to solve VMC with NQS.
First, we provide an initial overview of the general strategy to solve the optimisation problem at hand.
% We provide an initial overview of the optimisation problem at hand, and then put the problem in context with an overview on
% first-order optimisation strategies based on the standard gradient descent method.
Then, we provide the rationale and details for the specific optimisers we consider,
namely the Natural Gradient Descent and its KFAC approximation.
Finally, building on the shortcomings of KFAC, we discuss several improvements and test the newly developed optimisation strategies.
% Our discussion is relatively technical, but necessary in view of some of the shortcomings we experience in developing new optimisation strategies.
% This approach informed us in order to develop the new optimisation strategies presented in Section~\ref{sec:decisionalGradient}.

\subsection{General optimisation strategy}{\label{sec:general_optimisation}}
Solving a global, non-linear optimisation problem is a difficult task.
In this work, we consider optimisers that follow the same underlying strategy, sometimes
referred to as \emph{sequential quadratic programming}~\cite{nocedal2006numerical}.
In our case, the general problem takes the form,
\begin{subequations}
\begin{align}
    E^* &= \text{min}_{\theta} \ E(\theta) \ , \\
    \theta^* &= \text{argmin}_{\theta} \ E(\theta) \ .
\end{align}
\end{subequations}
This problem is divided into a sequence of simpler sub-problems which are solved iteratively.
The $n^{\text{th}}$ sub-problem consists of 
(a) an initial guess on the set of parameters, $\theta_n$; 
(b) a quadratic model, $M_n$, approximating the loss function
$E(\theta)$ locally; 
and (c) a trust region, $T_n$, specifying where the local model is expected
to be a good approximation to the loss function~\cite{nocedal2006numerical}.
Formally, solving the $n^{\text{th}}$ sub-problem consists in computing a new energy estimate, $E_{n+1}$,
at the updated parameters, $\theta_{n+1}$, namely
\begin{subequations}
\begin{align}
    E_{n+1} &= E(\theta_{n+1}) \ , \\
    \theta_{n+1} &= \theta_n + \text{argmin}_{\delta \in T_n} \ M_n(\delta) \ .
\end{align}
\end{subequations}
Here, $\delta$ represents the update in the parameters and the quadratic model $M_n$ takes the form
\begin{equation}
    M_n(\delta) = \frac{1}{2} \delta^\transpose Q \delta + L^\transpose \delta + C \ ,
\end{equation}
where $Q$ is a matrix; $L$, a vector; and $C$, a scalar.
In practice, the constrained quadratic optimisation sub-problem is replaced by
an equivalent unconstrained regularised sub-problem,
\begin{equation}
    \delta_n \equiv
    \text{argmin}_{\delta \in T_n} \ M_n(\delta)
    = \text{argmin}_{\delta} \left( M_n(\delta) + \frac{1}{2} \lambda_n \delta^\transpose R_n \delta \right) \ ,
    \label{eq:regularisedUpdate}
\end{equation}
where $\lambda_n$ is a damping parameter, characterising the size of the equivalent trust region
and $R_n$ is a symmetric positive semi-definite matrix describing the shape of the trust region.
This procedure is often referred to as Tikhonov regularisation~\cite{TikhonovBook1,nocedal2006numerical}.
Importantly, not only are the energy and the parameters, $(E_n,\theta_n)$,  updated at each iteration, 
but also the local quadratic model and its trust region (or regularisation), $(M_n(\delta), T_n)$.
Explicitly, Eq.~\eqref{eq:regularisedUpdate} has the general solution
\begin{equation}
    \delta_n = -\left( Q+\lambda_n R_n \right)^{-1} L \ .
\end{equation}
% For numerical reasons, sometimes it is useful to write the above equation as a linear system of equations to solve for $\delta$,
In practice $\delta_n$ is obtained by solving the linear system
\begin{equation}
\label{eq:optimisation_general_solution}
    \left( Q+\lambda_n R_n \right) \delta_n = -L \ .
\end{equation}

As an instructive example, the standard gradient descent method can be formulated in this framework.
In this case, one solves iteratively the locally constrained sub-problems associated to the model
\begin{align}
    M^{\text{GD}}_n(\delta) &= \nabla E(\theta_n)^\transpose \delta + E(\theta_n) \ , 
    \label{eq:gradientDescent1}
\end{align}
with trust region
\begin{align}
    \label{eq:gradientDescent2}
    T^{\text{GD}}_n &= \set{\delta : \norm{\delta}_2 \leq \alpha \norm{\nabla E(\theta_n)}_2} \ ,
\end{align}
where $\norm{.}_2$ denotes the standard Euclidean norm.
In this case, the update on the parameter $\delta_n$ reads explicitly
\begin{equation}
    \delta^{\text{GD}}_n = - \alpha \ \nabla E(\theta_n) \ .
\end{equation}
When solving this sequence of local constrained sub-problems, one is effectively performing a gradient descent
with a learning rate $\alpha$. 

\subsection{Natural Gradient Descent}\label{sec:NGD}

Equations~\eqref{eq:gradientDescent1}-\eqref{eq:gradientDescent2} indicate that 
the standard gradient descent algorithm relies on setting an isotropic
constraint on the size of the update parameter, $\delta_n$, based on the Euclidean norm.
In a general setting, the Euclidean norm has no reason to be an optimal choice.
Different gradient descent algorithms have been
designed in an attempt to outperform the standard version presented in the previous subsection. 
The \emph{natural gradient descent} (NGD) algorithm, which has been advocated to be optimal
for a class of optimisation problems, is of particular interest~\cite{amari1998natural,amari1998why}.
This algorithm is the starting point of the KFAC optimiser~\cite{martens2015optimizing}.

The NGD was originally defined in a supervised learning setting,
where the goal is to learn an empirically sampled distribution $q$ by minimising a loss function.
One particular loss function, chosen here for pedagogical reasons, is the cross-entropy, namely
\begin{equation}\label{eq:cross-entropy-loss}
    \mathcal{L}^{\text{sup}}(\theta) = \mathbb{E}_{X \sim q} \left[ - \ln p_{\theta}(X) \right] \ .
\end{equation}
Here,  $p_\theta$ is a probability distribution, depending on several parameters $\theta$, that is adjusted to 
minimise the cross-entropy loss.
In this case, the manifold of the probability distributions $p_{\theta}$ possesses a natural
geometry commonly referred to as \emph{information geometry}~\cite{amari1983foundation,amari2016information}.
The information geometry is defined by the Kullback-Leibler (KL) divergence between two probability distributions
\begin{equation}\label{eq:KLdiv}
    D_{\text{KL}}(p_{\theta} , \ p_{\theta'})
    \equiv \mathbb{E}_{X \sim p_{\theta}} \left[ - \ln p_{\theta'}(X)  - (- \ln p_{\theta}(X)) \right] \ .
\end{equation}
Although the KL divergence does not satisfy all the axioms of a distance, it defines locally a metric
obtained by Taylor-expanding it at second-order, namely
\begin{equation}
    D_{\text{KL}}(p_{\theta} , \ p_{\theta+\delta})
    = \frac{1}{2} \delta^\transpose F(\theta) \delta  + O\left(\delta^3\right)\ ,
\end{equation}
where $F(\theta)$ is the Fisher information matrix (FIM).
The expansion is performed around the “diagonal” point, $p_{\theta} = \ p_{\theta'}$.
From the expansion, one obtains an explicit expression for the FIM,
\begin{equation}
    F(\theta)_{\theta_i \theta_j}
    \equiv
    \mathbb{E}_{X \sim p_\theta} \left[ \partial_{\theta_i} \ln p_{\theta}(X) \partial_{\theta_j} \ln p_{\theta}(X)\right] \ ,
\end{equation}
where $\theta_i$ and $\theta_j$ are individual components of the global parameter vector, $\theta$.
In this context, the NGD consists simply in performing a gradient descent with a trust region 
defined by the norm associated to the FIM. For example, 
$
    T^{\text{NGD}}_n = \set{\delta : \delta^\transpose F(\theta_n) \delta \leq r^2} ,
$
where $r > 0$ denotes the maximum size of an update on $\delta_n$.
As mentioned earlier, it is more practical to work with an equivalent unconstrained regularised problem.
In the case of NGD for VMC, the regularised local model is
\begin{equation}
    M^{\text{NGD-reg}}_n(\delta)
    = \lambda_{F} \frac{1}{2} \delta^\transpose F(\theta_n) \delta + \nabla E(\theta_n)^\transpose \delta + E(\theta_n) \ ,
    \label{eq:NGDreg}
\end{equation}
where the damping $\lambda_F$ is a function of the size of $T^{\text{NGD}}_n$.
Equivalently, the update minimising $M^{\text{NGD-reg}}_n(\delta)$ can be expressed as
\begin{equation}
    \delta^{\text{NGD-reg}}_n = - \alpha \ F^{-1}(\theta_n) \nabla E(\theta_n) \ ,
        \label{eq:NGDreg_update}
\end{equation}
where $\alpha$ is a learning rate related to the size of $T^{\text{NGD}}_n$.

% In practice, NGD has been shown to be efficient for specific kinds
% of supervised learning problems~\cite{amari1998natural,amari1998why}.
The reasons for using NGD as an optimiser for our NQS are diverse.
In the context of VMC, a quantum generalisation of NGD, the so-called
Stochastic Reconfiguration (SR),
provides the foundation for many state-of-the-art optimisers
used with NQS~\cite{Sorella2005,Stokes2020,rigo2023solving}.
SR has been shown to converge to exact results in challenging many-body settings~\cite{chen2023efficient,rende2023simple}.
% has been observed to perform well
% and provides the foundation for many state-of-the-art optimisers
% used with NQS~\cite{Sorella2005,Stokes2020,rigo2023solving}.
The metric employed in SR is the real part of the so-called
quantum geometric tensor (QGT)~\cite{Stokes2020}.
While SR and NGD differ in general, the QGT can be shown to reduce to the FIM when considering
a manifold of real-valued ansätze which is the case of the NQS we consider here.
% This is due to the fact that the phase of the wave function ansatz is constant,
% as a function of the parameters of the NQS, in a small enough neighbourhood around $\theta$.
For VMC calculations, the use of SR is motivated by the fact that the QGT favours directions
with a higher signal-to-noise ratio in the estimator of the energy gradient,
and by its interpretation as an imaginary-time projection technique~\cite{Sorella2005}.
Beyond the sole case of VMC calculations, two critical properties of the FIM have been
raised for explaining the performances of NGD for general supervised learning problems:
% The use of NGD and the associated FIM goes beyond the sole case of VMC calculations.
% % While there is clearly some physical motivation behind SR, the exact reasons for the success of the NGD are still under study~\cite{martens2020new}.
% For general supervised learning problems, two critical properties of the FIM have been
% raised for explaining the performances of NGD:
(i) the FIM is symmetric definite positive and (ii) it provides an approximation to
the Hessian of the loss function~\cite{martens2020new}.
Point (i) is important to avoid instabilities in the directions associated to negative eigenvalues
of the quadratic term, so that updates in the parameters remain bounded.
Point (ii) leads to a re-interpretation of NGD as a second-order optimisation scheme, where the Hessian of the loss
function is approximated by the FIM. This motivates the minimisation of a model
\begin{equation}
    M^{\text{NGD}}_n(\delta)
    = \frac{1}{2} \delta^\transpose F(\theta_n) \delta + \nabla E(\theta_n)^\transpose \delta + E(\theta_n) \ ,
\end{equation}
which leads to the update in the parameters
\begin{equation}\label{eq:NGD-update}
    \delta^{\text{NGD}}_n = - F^{-1}(\theta_n) \nabla E(\theta_n) \ .
\end{equation}
This interpretation of NGD is equivalent to fixing the learning rate to unity, $\alpha=1$, in Eq.~\eqref{eq:NGDreg_update}.

\subsection{KFAC algorithm}\label{sec:KFAC_algorithm}

A practical implementation of the NGD algorithm requires computing the update in Eq.~\eqref{eq:NGD-update}.
In the case of NN models, the number of parameters, $N$, is large ($N \gg 10^4$).
Computing exactly the update $\delta^{\text{NGD}}_n$ at every epoch can be computationally prohibitive.
The goal of the KFAC algorithm is to compute an approximation to the update
$\delta^{\text{NGD}}_n$ with a time complexity of $O\left(N\right)$
while keeping the number of epochs to convergence roughly constant. We note that,
for the specific case of NQS, 
%there are alternative proposals to implement this update calculations employing fast Jacobian-vector product algorithms~\cite{rende2023simple,chen2023efficient,vicentini2022netket}.
it has been recently shown that one can reformulate the update $\delta^{\text{NGD}}_n$ in a way which significantly reduces the time complexity of the implementation~\cite{rende2023simple,chen2023efficient,vicentini2022netket}.

Consider a feed-forward NN of arbitrary depth. 
For a given layer $l$, we denote the  
weight matrix $W^{(l)}$ %$ \in \mathbb{R}^{H \times H}$ 
and, for simplicity, we assume it to be of size $H \times H$ and without bias. 
The input row-vector $f^{(l-1)} \in \mathbb{R}^{1 \times H}$ and the output row-vector $h^{(l)} \in \mathbb{R}^{1 \times H}$
are related according to
$ %\begin{equation}
    h^{(l)} = \tanh \left( f^{(l-1)} W^{(l)\transpose} \right) \ , 
$ %\end{equation}
where the activation function $\tanh$ is applied element-wise\footnote{We focus on this activation function because it is the one of choice
in our NQS model~\cite{keeble2023machine}, but any other choice would not change our conclusions. 
}. In this work, we follow the convention used in the PyTorch framework,
where inputs and outputs of a layer are row vectors.

In the KFAC algorithm originally described in Ref.~\cite{martens2015optimizing},
the FIM is approximated in two ways. First, the matrix elements between parameters of different layers
are neglected. The approximated FIM, $\breve{F}(\theta)$, is block-diagonal and reads
\begin{equation}
    \breve{F}(\theta)_{W^{(l)}_{ij} W^{(l')}_{i'j'}}
    \equiv
    \begin{cases}
        F(\theta)_{W^{(l)}_{ij} W^{(l')}_{i'j'}} & \text{ if } l = l' \\
        0 & \text{ if } l \neq l' \ ,
    \end{cases}
\end{equation}
where $W^{(l)}_{ij}$ is the parameter associated to the matrix element $(i,j)$
of the $l^{\text{th}}$ layer weight matrix. 
Second, each layer-dependent block is approximated by a Kronecker product.
Using the chain rule, the  FIM derivatives can be rewritten as
\begin{align}
    \frac{\partial \ln p_{\theta}}{\partial {W^{(l)}_{ij}}}
    &= \frac{\partial h^{(l)}_{1i}}{\partial {W^{(l)}_{ij}}}
        \frac{\partial \ln p_{\theta}}{\partial h^{(l)}_{1i}} 
    =  f^{(l-1)}_{1j}
        \times
        \tanh'\left( \sum_{k} f^{(l-1)}_{1k} W^{(l)}_{ik} \right)
        \times \frac{\partial \ln p_{\theta}}{\partial h^{(l)}_{1i}} \ .
\end{align}
Introducing the so-called forward activations, $a^{(l-1)}$, and the backward sensitivities, $e^{(l)}$, as the row vectors
\begin{subequations}
\begin{align}
    a^{(l-1)}_{1j} &\equiv f^{(l-1)}_{1j} \ , \\
    e^{(l)}_{1i} &\equiv
        \tanh'\left( \sum_{k} f^{(l-1)}_{1k} W^{(l)}_{ki} \right)
        \times \frac{\partial \ln p_{\theta}}{\partial h^{(l)}_{1i}} \ ,
\end{align}
\end{subequations}
we obtain
\begin{equation}\label{eq:KroneckerFactorisationDerivative}
    \text{vec}\left(\frac{\partial \ln p_{\theta}}{\partial W^{(l)}}\right)
    =
    \text{vec}\left( e^{(l)} a^{(l-1)\transpose} \right)
    = a^{(l-1)} \otimes e^{(l)} \ ,
\end{equation}
where $\text{vec}(.)$ and $(.\otimes.)$ denote respectively the standard vectorisation (stacking columns of a matrix)
and Kronecker product operations.
Using Eq.~\eqref{eq:KroneckerFactorisationDerivative}, the $l^{\text{th}}$ block of the FIM can be rewritten
in vectorised form as
\begin{align}
    \breve{F}^{(l)}(\theta)
    &=
    \mathbb{E}_{p_\theta}
    \left[
        \text{vec}\left(\frac{\partial \ln p_{\theta}}{\partial W^{(l)}}\right)
        \text{vec}\left(\frac{\partial \ln p_{\theta}}{\partial W^{(l)}}\right)^\transpose
    \right]
    =
    \mathbb{E}_{p_\theta}
    \left[
        \left(a^{(l-1)} \otimes e^{(l)}\right)
        \left(a^{(l-1)} \otimes e^{(l)}\right)^\transpose
    \right] \nonumber \\
    &=
    \mathbb{E}_{p_\theta}
    \left[
        \left(a^{(l-1)} a^{(l-1)\transpose}\right)
        \otimes
        \left(e^{(l)} e^{(l)\transpose}\right)
    \right] \ . \label{eq:fisher_block}
\end{align}
Finally, the Kronecker-factored approximation on the $l^{\text{th}}$ block of the FIM,
$\breve{F}^{(l)}_{\text{KFAC}}(\theta)$, is obtained by exchanging the order of 
the Kronecker product and the statistical average, i.e.\ 
\begin{equation}\label{eq:KFAC-approximation}
    \breve{F}^{(l)}_{\text{KFAC}}(\theta)
    \equiv
    \mathbb{E}_{p_\theta}\left[a^{(l-1)} a^{(l-1)\transpose}\right]
    \otimes
    \mathbb{E}_{p_\theta}\left[e^{(l)} e^{(l)\transpose}\right] \ .
\end{equation}
The resulting block-diagonal KFAC FIM will be denoted as $\breve{F}_{\text{KFAC}}(\theta)$ from now on. 

Assuming a unit learning rate, the update in the parameters when using the KFAC approximation thus reads
\begin{equation}\label{eq:KFAC-direction-update}
    \Delta^{\text{KFAC}}_n \equiv - \breve{F}^{-1}_{\text{KFAC}}(\theta_n) \nabla E(\theta_n) \ .
\end{equation}
We employ a capital $\Delta^{\text{KFAC}}_n$ to differentiate it from previous parameter updates. 
The main advantage of the KFAC approximation is that, instead of having to invert
a matrix of dimension $(L \times H^2) \times (L \times H^2)$, one can invert separately the $2L$
Kronecker factors in Eq.~\eqref{eq:KFAC-approximation}, which are much smaller matrices of dimension $H \times H$.
The main drawback is that the KFAC approximation is very crude and can only catch the very coarse structure of the
exact FIM~\cite{martens2015optimizing}.
For completeness, let us also mention that the two Kronecker
factors are averaged over successive epochs, with a maximum exponentially decay value set to $0.9$~\cite{martens2015optimizing}.

To compensate for the poor quality of the update $\Delta^{\text{KFAC}}_n$, a second step
was designed in the original KFAC algorithm~\cite{martens2015optimizing}. The idea is that, whenever $\Delta^{\text{KFAC}}_n$
provides a good direction, we would like to take a big step. In contrast, when the direction
is poor, we would like to take a small step, to avoid increasing the loss function.
In the original KFAC algorithm, this is achieved by computing an optimal rescaling factor $\alpha$
such that the quadratic model $M^{\text{NGD}}_n \left( \alpha \Delta^{\text{KFAC}}_n \right)$, which uses the \emph{exact} FIM,
is minimised. In addition, a momentum in the previous update $\delta^{\text{KFAC}}_{n-1}$ is usually included.
A natural way to include both improvements consists in minimising 
$M^{\text{NGD}}_n\left (\alpha \Delta^{\text{KFAC}}_n + \mu \delta^{\text{KFAC}}_{n-1} \right)$
over the two parameters $\alpha$ and $\mu$. Explicitly, the optimal coefficients $\alpha^*$ and $\mu^*$ are obtained from the
linear system
\begin{equation}\label{eq:rescaling-momentum-opt}
    \begin{pmatrix}
        \alpha^* \\
        \mu^*
    \end{pmatrix}
    =
    -
    \begin{pmatrix}
        \Delta^\transpose_n F(\theta_n) \Delta_n & \Delta^\transpose_n F(\theta_n) \delta_{n-1} \\
        \delta^\transpose_{n-1} F(\theta_n) \Delta_n & \delta^\transpose_{n-1} F(\theta_n) \delta_{n-1}
    \end{pmatrix}^{-1}
    \begin{pmatrix}
        \nabla E(\theta_n)^\transpose \Delta_n \\
        \nabla E(\theta_n)^\transpose \delta_{n-1}
    \end{pmatrix} \ ,
\end{equation}
where we momentarily drop the superindex KFAC in all the $\Delta_n$ and $\delta_n$ 
updates for simplicity.
Using the \emph{exact} FIM is crucial to help correct cases where the KFAC approximation is too crude.
This correction remains relatively cheap, since Eq.~\eqref{eq:rescaling-momentum-opt} only requires computing
a couple of matrix-vector calculations.

Last but not least, the original derivation of KFAC 
describes a particular Tikhonov regularisation
in order to stabilise the overall algorithm~\cite{martens2015optimizing}.
Such a regularisation is of critical importance, since it limits how large the update in the parameters can be.
If the trust region is too small, the benefits of using a second-order optimiser will be null.
If the trust region is too large, instabilities will kick in and ruin any chances of reaching convergence.

The KFAC FIM is used to estimate the direction of the update,
while the exact FIM is used to evaluate an optimal rescaling of the direction.
Consequently, two distinct trust regions are considered, which reflect the different level of trust one has
in the direction, $\Delta^{\text{KFAC}}_n$, and the rescaling factor, $\alpha^*$.
Those are implemented by introducing two damping factors, $\gamma_n$ and $\lambda_n$, which regularise
$\breve{F}_{\text{KFAC}}(\theta_n)$ and $F(\theta_n)$, respectively.
More explicitly, the regularised matrices are obtained from the substitutions
\begin{subequations}
\begin{align}
    \breve{F}^{(l)}_{\text{KFAC}}(\theta_n)
        &\to
        \left(\mathbb{E}_{p_{\theta_n}}\left[a^{(l-1)} a^{(l-1)\transpose}\right] + \pi^{(l)} \gamma_n \ I \right)
        \otimes
        \left(\mathbb{E}_{p_{\theta_n}}\left[e^{(l)} e^{(l)\transpose}\right] + \frac{\gamma_n}{\pi^{(l)}} \ I\right) \ , \\
    F(\theta_n)
        &\to F(\theta_n) + \lambda_n \ I \ ,
\end{align}
\end{subequations}
where $\pi^{(l)}$ is a factor that is automatically adjusted to minimise the cross-term obtained
when expanding the Kronecker product~\cite{martens2015optimizing}. 
In practice, the damping factors are dynamically adapted every $T$ epochs.

\begin{figure*}[t]
\begin{center}
\includegraphics[width=0.7\linewidth, trim=0 130 0 130]{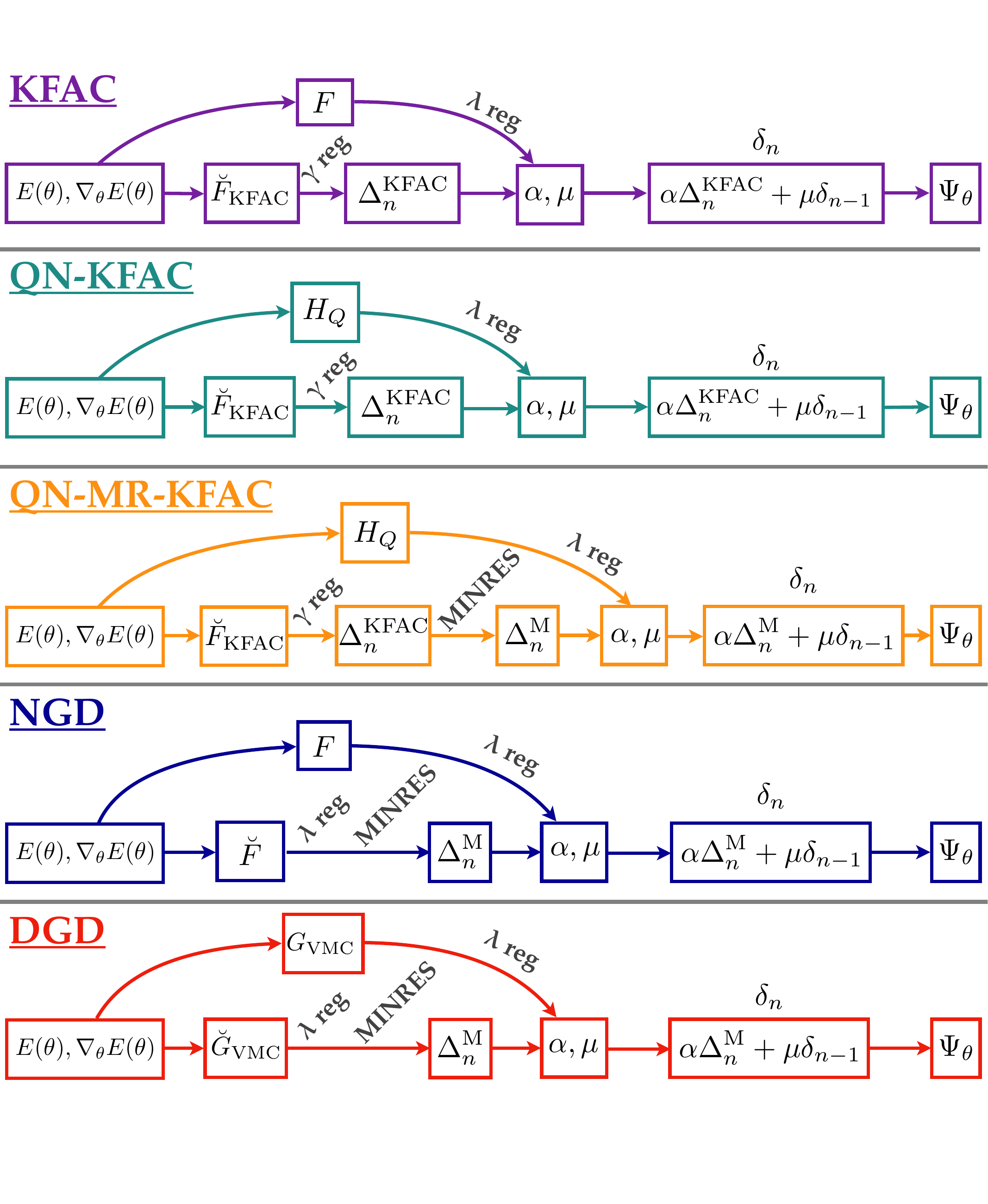}
\end{center}
\caption{Schematic view of the algorithms associated to the optimisers discussed in this work.}
\label{fig:masterFigureoptimisers}
\end{figure*}

While we try to be as close to the original KFAC algorithm as possible, our numerical implementation for the NQS toy model
slightly differs in the dynamical adjustment of the damping factors.
The damping factor $\lambda_n$ is adapted following a Levenberg-Marquardt rule~\cite{more2006levenberg}, namely
\begin{equation}\label{eq:LevenbergMarquardt}
    \begin{cases}
        \rho_n < 0.25, & \lambda_{n+1} = \frac{\lambda_{n}}{\omega_1} \\
        0.25 \leq \rho_n \leq 0.75, & \lambda_{n+1} = \lambda_{n} \\
        \rho_n > 0.75, & \lambda_{n+1} = \omega_1 \ \lambda_{n}
    \end{cases} \ .
\end{equation}
The reduction ratio,
\begin{equation}\label{eq:reductionRatio}
    \rho_n \equiv \frac{E \left(\theta_{n} + \delta_n^{\text{KFAC}} \right) - E(\theta_{n})}{M^{\text{NGD}}_n \left( \delta_n^{\text{KFAC}} \right) - M^{\text{NGD}}_n(0)} \ ,
\end{equation}
measures the discrepancy between the actual variation of the loss (numerator) and the anticipated one from the quadratic model (denominator). 
Unlike in the original KFAC implementation, however, we also account for fluctuation on the energy estimation by introducing a tolerance
on increasing the energy, i.e.\ if $ E(\theta) \leq E(\theta+\delta) \leq E(\theta) + \epsilon$,
we default to the case where the damping $\lambda_n$ is reduced.
In all our calculations $\epsilon$ is set to $10\%$ of $E(\theta)$. 
The damping $\gamma_n$ is adapted with a greedy algorithm by testing which damping among
$(\gamma_n, \omega_2 \ \gamma_n, \frac{\gamma_n}{\omega_2})$ leads to the best update.
For all the calculations performed in this work, we set $T=5$,
$\omega_1 = ({19}/{20})^T$, $\omega_2 = \sqrt{{19}/{20}}^T$ and
the initial damping values are set to $\lambda_0 = 10^3$ and $\gamma_0 = \sqrt{10^3}$.
For clarity, we provide a schematic overview of the KFAC algorithm in Fig.~\ref{fig:masterFigureoptimisers}.
For more details, we refer the reader to Ref.~\cite{martens2015optimizing}. 

\subsection{Improving the scaling estimation}\label{sec:scaling_improvement}
In its inception, the KFAC optimiser was tested on several deep autoencoder
optimisation problems with standard datasets (MNIST, CURVES and FACES), which are commonly used
to benchmark NN optimisation methods~\cite{hinton2006reducing,martens2010deep}.
All those cases fall in the category of supervised learning for standard feed-forward NNs.
Beyond such standard cases, the original KFAC optimiser has to be adapted.
For instance, if convolution layers are a part of the NN architecture, 
Refs.~\cite{grosse2016kronecker,ba2017distributed} suggested modifications to the KFAC
approximation of the FIM.
In the case of VMC with NQS, several attempts have been put forward to exploit the
benefits of KFAC~\cite{pfau2020ab,wilson2021simulations,wilson2022wave}.
To bypass instabilities encountered when applying the standard KFAC optimiser,
several \emph{ad hoc} adjustments have been implemented to rectify the update
on the parameters~\cite{pfau2020ab}.
In particular, Ref.~\cite{pfau2020ab} describes the use of a scheduled learning rate,
a constant damping, a norm constraint  and the discarding of any momentum in the parameter updates.
Those adjustments are meant to replace the rescaling phase using the exact FIM
and the dynamically adjusted damping parameters advocated in Ref.~\cite{martens2015optimizing}.
Apart from the lack of theoretical motivation for such adjustments,
it is consistently reported within the literature that heuristics
and fine-tuning are required to obtain good
performance~\cite{pfau2020ab,wilson2021simulations,wilson2022wave}.
In this section, we advocate for a different approach, which avoids any such %fiddling 
heuristic adjustment of hyperparameters,
while remaining close to the original philosophy of Ref.~\cite{martens2015optimizing}.
We start by providing an overview of the performance associated to our implementation of KFAC for the
toy model  described in Sec.~\ref{sec:quantumManyBody}, before suggesting improvements to the
rescaling part of the algorithm.

\begin{figure*}
\includegraphics[width=0.94\linewidth, trim=180 80 180 0]
{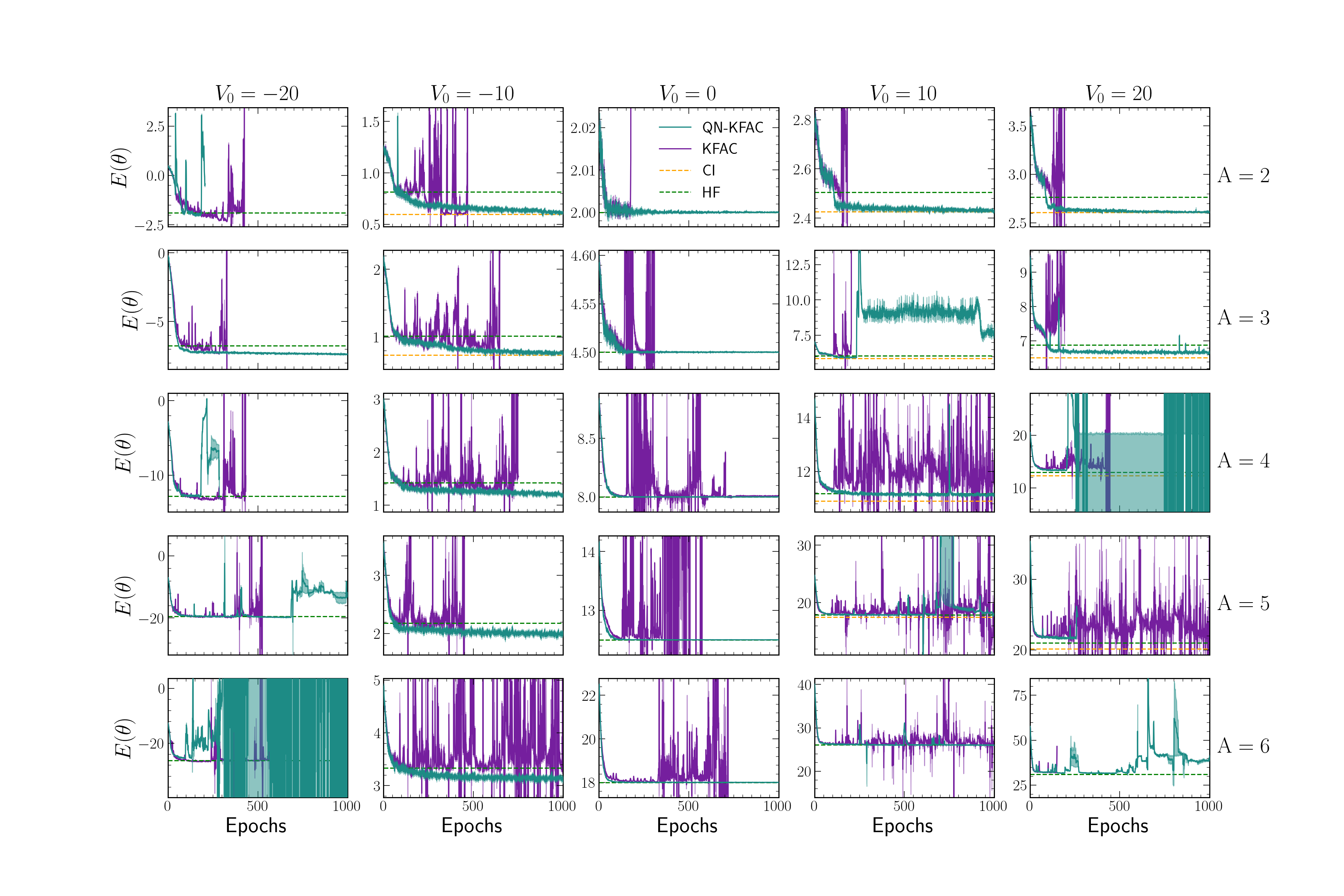}
	\caption{Evolution of the energy $E(\theta)$ as a function of the number of epochs for the KFAC
    (purple solid line) and for the QN-KFAC (green solid line) optimisers. 
    The corresponding shaded area represents the standard deviation of the Monte Carlo energy estimate.
     From top to bottom, the panels correspond to results for $A=2$, $3$, $4$, $5$ and $6$ particles, respectively.
    From left to right, $V_0$ goes from $-20$ to $20$ in increments of $10$ units.
    HF and CI results
    are displayed for reference, with horizontal green and orange dashed lines.
    CI results are displayed for $A \leq 5$ and $V_0 \geq -10$ where we have access to
    converged results~\cite{keeble2023machine}.
}
	\label{fig:KFAC_vs_QN-KFAC}
\end{figure*}

To compare the performance of the different optimisers,
we show the evolution of the energy as a function of the number of epochs in the panels of Fig.~\ref{fig:KFAC_vs_QN-KFAC}. 
From top to bottom, the different rows correspond to increasing number of particles, from $A=2$ in the top 
to $A=6$ in the bottom row. From left to right, the results scan the interaction strength in a range 
$V_0 \in \set{-20, -10, 0, 10, 20}$. Dashed lines provide an indication of the expected results for the HF approximation 
and for the CI case.
The optimisation process in all cases is stopped after $1000$ epochs.
Very few runs reach this final number and most lead to numerically
unstable results much earlier than that. 
As in previous KFAC implementations\cite{pfau2020ab,wilson2021simulations,wilson2022wave}, we confirm that the performance 
of the KFAC optimiser is indeed fluctuating.
In many cases, the energy decreases steadily within the first $50$ to $100$ epochs,
only to become extremely erratic after that point. 
None of the KFAC runs manage to converge to the benchmark values, regardless of the number
of particles and of the strength of the interaction. 
We take this as a confirmation that the KFAC algorithm, in its original formulation, is unfit for VMC with NQS.

% \begin{figure*}
% \includegraphics[width=0.94\linewidth, trim=180 80 180 0]
% {figures/rasterized_figures/convergence_KFAC_vs_QN-KFAC.png}
% 	\caption{Evolution of the energy $E(\theta)$ as a function of the number of epochs for the KFAC
%     (purple solid line) and for the QN-KFAC (green solid line) optimisers. 
%     The corresponding shaded area represents the standard deviation of the Monte Carlo energy estimate.
%      From top to bottom, the panels correspond to results for $A=2$, $3$, $4$, $5$ and $6$ particles, respectively.
%     From left to right, $V_0$ goes from $-20$ to $20$ in increments of $10$ units.
%     HF and CI results
%     are displayed for reference, with horizontal green and orange dashed lines.
%     CI results are displayed for $A \leq 5$ and $V_0 \geq -10$ where we have access to
%     converged results~\cite{keeble2023machine}.
% }
% 	\label{fig:KFAC_vs_QN-KFAC}
% \end{figure*}

% One option to compensate the strong oscillations is to try and improve the rescaling phase of the optimiser.
% To improve the KFAC algorithm and reduce the strong oscillations we first focus on improving
% its rescaling phase.

In order to restore the validity of KFAC in VMC calculations, we first focus on improving
the rescaling part of the optimiser.
The main argument put forward in favour of using $M^{\text{NGD}}_n$ in the rescaling phase
is that the FIM can be interpreted as an approximation to the Hessian of the loss
function~\cite{martens2015optimizing,martens2020new}.
However, as emphasised in Refs.~\cite{martens2015optimizing,martens2020new}, the validity of this argument
relies critically on the assumption that one is trying to solve a \emph{supervised learning} problem.
In this case, the loss function quantifies how far the probability distribution $p_\theta$, modelled
by the NN, is from the target distribution, $q$. An example of such loss function is the
cross-entropy, given in Eq.~\eqref{eq:cross-entropy-loss}.
In VMC, however, one is solving a \emph{reinforcement learning} problem, which is fundamentally different.
% In fact, in VMC one does not have, a priori, any information about the target wave function or energy; the loss function uses information only about the current prediction to improve the ansatz.
In this case, the loss function we use is the energy $E(\theta)$ which
characterises the quality of the NQS independently of the knowledge of the target wave function.
As a consequence, the underlying rationale justifying the use of the FIM
as an approximate Hessian of the loss function is lost.

% In order to restore the validity of KFAC in VMC calculations, we replace the exact FIM
To mitigate this issue, we replace the exact FIM
used in the rescaling phase by a quasi-Hessian matrix, $H_Q(\theta)$, whose role is to provide a better quadratic
model than $M^{\text{NGD}}_n$, namely
\begin{equation}
    M^{\text{QN}}_n(\delta)
    \equiv
    \frac{1}{2} \delta^\transpose H_Q(\theta_n) \delta + \nabla E(\theta_n)^\transpose \delta + E(\theta_n) \ .
\end{equation}
Let us emphasise that we still use the KFAC FIM in the estimation of the direction as in KFAC.
Throughout this work, we will refer to the resulting optimisation algorithm as \emph{quasi-Newton} KFAC (QN-KFAC),
and we employ the superscript $\text{QN}$ in the corresponding quantities.

The remaining question consists in defining which quasi-Hessian matrix $H_Q(\theta)$ to use.
In the case of supervised learning, the performance of NGD has been connected
to the similarity between the FIM $F(\theta)$ and the Generalised-Gauss-Newton (GGN) matrix
$G(\theta)$~\cite{martens2020new}. In particular, it has been shown in
Refs.~\cite{heskes2000natural,pascanu2014revisiting} that for several standard loss functions
(squared error, cross-entropy, softmax, etc) we have $F(\theta)=G(\theta)$. This is crucial since
the GGN matrix is obtained by neglecting the Hessian of the output of the NN
in the calculation of the Hessian of the loss function so that the FIM can be interpreted as
an approximation of the Hessian of the cost function~\cite{martens2020new}.
Therefore, to improve the rescaling phase of the algorithm, we follow a similar path
and define the quasi-Hessian $H_Q(\theta)$ to be the matrix obtained from the Hessian of the energy
$E(\theta)$, after neglecting the Hessian of $\ln\abs{\Psi_\theta}$.
More specifically, the Hessian of the energy reads~\cite{lin2000optimisation}
\begin{align}
    \partial_{\theta_i}\partial_{\theta_j} E(\theta)
    &=
    2 \mathbb{E}_{X\sim p_\theta}
    \left[
        \left(
            E_{L,\theta} (X)
            - E(\theta)
        \right)
    \right. \nonumber %\\
    \left.
        \times
        \left(
            \partial_{\theta_i} \partial_{\theta_j}\ln\abs{\Psi_{\theta}(X)}
            - \mathbb{E}_{X\sim p_\theta} \left[ \partial_{\theta_i} \partial_{\theta_j}\ln\abs{\Psi_{\theta}(X)} \right]
        \right)
    \right] \nonumber \\
    &\phantom{=}
    + 2 \mathbb{E}_{X\sim p_\theta}
    \left[
        \left(
            \partial_{\theta_i} E_{L,\theta} (X)
            - \mathbb{E}_{X\sim p_\theta} \left[ \partial_{\theta_i} E_{L,\theta} (X) \right]
        \right)
    \right. \nonumber \\
    &\phantom{= + 2 \mathbb{E}_{X\sim p_\theta} \ [ \ }
    \left.
        \times
        \left(
            \partial_{\theta_j}\ln\abs{\Psi_{\theta}(X)}
            - \mathbb{E}_{X\sim p_\theta} \left[ \partial_{\theta_j}\ln\abs{\Psi_{\theta}(X)} \right]
        \right)
    \right] \nonumber \\
    &\phantom{=}
    + 4 \mathbb{E}_{X\sim p_\theta}
    \left[
        \left(
            E_{L,\theta} (X)
            - E(\theta)
        \right)
    \right. \nonumber \\
    &\phantom{= + 4 \mathbb{E}_{X\sim p_\theta} \ [ \ }
    \left.
        \times
        \left(
            \partial_{\theta_i}\ln\abs{\Psi_{\theta}(X)}
            - \mathbb{E}_{X\sim p_\theta} \left[ \partial_{\theta_i}\ln\abs{\Psi_{\theta}(X)} \right]
        \right)
    \right. \nonumber \\
    &\phantom{= + 4 \mathbb{E}_{X\sim p_\theta} \ [ \ }
    \left.
        \times
        \left(
            \partial_{\theta_j}\ln\abs{\Psi_{\theta}(X)}
            - \mathbb{E}_{X\sim p_\theta} \left[ \partial_{\theta_j}\ln\abs{\Psi_{\theta}(X)} \right]
        \right)
    \right]  \ . \label{eq:hessian_energy}
\end{align}
After setting $\partial_{\theta_i} \partial_{\theta_j}\ln\abs{\Psi_{\theta}(X)}$ to zero,
the induced quasi-Hessian reads
\begin{align}
    H_Q(\theta)_{\theta_i\theta_j}
    &=
    2 \mathbb{E}_{X\sim p_\theta}
    \left[
        \left(
            \partial_{\theta_i} E_{L,\theta} (X)
            - \mathbb{E}_{X\sim p_\theta} \left[ \partial_{\theta_i} E_{L,\theta} (X) \right]
        \right)
    \right. \nonumber \\
    &\phantom{= 2 \mathbb{E}_{X\sim p_\theta} \ [ \ }
    \left.
        \times
        \left(
            \partial_{\theta_j}\ln\abs{\Psi_{\theta}(X)}
            - \mathbb{E}_{X\sim p_\theta} \left[ \partial_{\theta_j}\ln\abs{\Psi_{\theta}(X)} \right]
        \right)
    \right] \nonumber \\
    &\phantom{=}
    + 4 \mathbb{E}_{X\sim p_\theta}
    \left[
        \left(
            E_{L,\theta} (X)
            - E(\theta)
        \right)
    \right. \nonumber %\\
    \left.
        \times
        \left(
            \partial_{\theta_i}\ln\abs{\Psi_{\theta}(X)}
            - \mathbb{E}_{X\sim p_\theta} \left[ \partial_{\theta_i}\ln\abs{\Psi_{\theta}(X)} \right]
        \right)
    \right. \nonumber \\
    &\phantom{= + 4 \mathbb{E}_{X\sim p_\theta} \ [ \ }
    \left.
        \times
        \left(
            \partial_{\theta_j}\ln\abs{\Psi_{\theta}(X)}
            - \mathbb{E}_{X\sim p_\theta} \left[ \partial_{\theta_j}\ln\abs{\Psi_{\theta}(X)} \right]
        \right)
    \right]  \ . \label{eq:quasi-hessian}
\end{align}
In practice, to reduce the statistical noise of the estimator of the quasi-Hessian
we use the symmetrised estimator introduced in Ref.~\cite{umrigar2005energy}.  
A key advantage of using a quasi-Hessian over the full Hessian comes from its
structure as a statistical average of outer products. This structure allows to
conserve a numerical complexity linear in the number of parameters of the NQS,
like for the FIM~\cite{schraudolph2002fast}.
However, a critical drawback is that the $H_Q(\theta)$ is not in general positive semi-definite,
which can lead to instabilities if the damping $\lambda_n$ is not large enough.
In addition, the formula given in Eq.~\eqref{eq:quasi-hessian}
requires computing third-order derivatives of the wave function.
These arise from the second-order derivatives with respect to the input
in the kinetic energy term of the local energy, and the derivatives with respect to the parameters.
These derivatives should be handled with care in order to avoid numerical 
instabilities~\cite{pfau2020ab}. 
A schematic overview of the QN-KFAC algorithm is provided in Fig.~\ref{fig:masterFigureoptimisers}.

We now proceed to compare the VMC energy evolution employing  QN-KFAC and KFAC.
We stress that we perform systematic calculations
with all hyperparameters of the two optimisers kept fixed to the aforementioned values (see also the Supplemental Material). This minimises the differences in the 
optimisation scheme and provides a general overview in different 
physical conditions. 
We observe that QN-KFAC (green lines) leads to a dramatic increase in overall stability. 
We find that for a small number of particles, $A\leq3$, QN-KFAC almost always manages to converge to
reasonable values. For larger particle numbers, the convergence pattern is more erratic. In some cases, we
notice strong oscillations. In others, we observe jumps in energy predictions. The best-behaved convergence patterns
are found for $V_0=0$ and $V_0=-10$. 
While this improvement is encouraging, we also notice that, even in the best cases, the late stage
convergence can be slow. For instance, if we focus on the second column, corresponding to $V_0=-10$, we observe that,
after a quick convergence during the first $100$ epochs, the QN-KFAC optimiser seems to struggle to converge
fully before reaching our stopping criterion of $1000$ epochs.
Comparing QN-KFAC with CI in the $A=2$ case (top panel), it is clear that QN-KFAC
converged to the exact result. However, for $A=6$ particles and the same interaction strength,
in the absence of exact CI benchmarks, it is not clear whether or not the QN-KFAC finished converging. 

The QN-KFAC optimiser provides a first improvement over KFAC in our NQS toy model simulations.
This was achieved by modifying the quadratic model used in the rescaling phase of the KFAC algorithm.
In order to tackle the remaining issues observed when testing QN-KFAC, like the presence of strong oscillations
in some cases and the slow convergence pattern in the late stage of the optimisation process, we now proceed to
improve on the direction estimation phase of the optimiser.

\subsection{Improving the direction estimation}\label{sec:direction_improvement}

The original KFAC algorithm was developed for supervised learning problems with standard feed-forward NNs.
In the case of a reinforcement learning problem like VMC, 
the theoretical motivation for using the FIM is lost.
This prompted the changes in the rescaling phase of the optimiser,
as discussed in the previous subsection.
In this subsection, we further argue that using an NQS architecture 
additionally breaks the original rationale motivating the Kronecker factorisation,
which lies at the heart of the KFAC algorithm~\cite{martens2015optimizing}.

As mentioned in section~\ref{sec:KFAC_algorithm}, the direction of the update in the parameters
$\delta^{\text{KFAC}}_n$ is obtained based on two approximations of the complete FIM:
(i) the FIM is assumed to be block-diagonal in the layers and
(ii) each block is further approximated by exchanging
the Kronecker product and the statistical average, thus replacing Eq.~\eqref{eq:fisher_block} by
Eq.~\eqref{eq:KFAC-approximation}. The latter approximation
is discussed at length in the original derivation of KFAC in Ref.~\cite{martens2015optimizing}. 
The validity of Eq.~\eqref{eq:fisher_block}, however, is unclear.
The chain rule used to derive it relies on the specific NN architecture.
In the case of a standard feed-forward NN,
% considered in Ref.~\cite{martens2015optimizing},
Eq.~\eqref{eq:fisher_block} does hold.
% as discussed in section~\ref{sec:KFAC_algorithm}.
% Equation~\eqref{eq:fisher_block}
However, it breaks down for other types of NN.
For example, this is the case for convolution layers where the original KFAC algorithm
had to be adapted~\cite{grosse2016kronecker}.

In the case of NQS for fermionic systems,
permutation-equivariant layers also break Eq.~\eqref{eq:fisher_block}.
To understand why, let us consider the $l^{\text{th}}$ permutation-equivariant layer with a shared
weight matrix $W^{(l)}\in\mathbb{R}^{H \times H}$ with input $f^{(l-1)} \in \mathbb{R}^{A \times H}$
and output $h^{(l)} \in \mathbb{R}^{A \times H}$, related through the activation function
$ %\begin{equation}
    h^{(l)} = \tanh \left( f^{(l-1)} W^{(l)\transpose} \right) .
$ %\end{equation}
Using the chain rule, the derivatives in the FIM for a fermionic NN now read
\begin{align}\label{eq:multidim_chain_rule}
    \frac{\partial \ln p_{\theta}}{\partial {W^{(l)}_{ij}}}
    &= \sum^{A}_{m=1} \frac{\partial h^{(l)}_{mi}}{\partial {W^{(l)}_{ij}}}
        \frac{\partial \ln p_{\theta}}{\partial h^{(l)}_{mi}} 
    =  \sum^{A}_{m=1} f^{(l-1)}_{mj}
        \times
        \tanh'\left( \sum_{k} f^{(l-1)}_{mk} W^{(l)}_{ik} \right)
        \times \frac{\partial \ln p_{\theta}}{\partial h^{(l)}_{mi}} \ .
\end{align}
In this case, the forward activities $a^{(l-1)}$ and backward sensitivities $e^{(l)}$
are not row vectors, but rather matrices
\begin{subequations}
\begin{align}
    a^{(l-1)}_{mj} &\equiv f^{(l-1)}_{mj} \ , \\
    e^{(l)}_{mi} &\equiv
        \tanh'\left( \sum_{k} f^{(l-1)}_{mk} W^{(l)}_{ki} \right)
        \times \frac{\partial \ln p_{\theta}}{\partial h^{(l)}_{mi}} \ .
\end{align}
\end{subequations}
With this, we recover formally the same result as Eq.~\eqref{eq:KroneckerFactorisationDerivative},
\begin{equation}\label{eq:FFN_FactorisationDerivative}
    \text{vec}\left(\frac{\partial \ln p_{\theta}}{\partial W^{(l)}}\right)
    =
    \text{vec}\left( e^{(l)} a^{(l-1)\transpose} \right) \ .
\end{equation}
However, because the forward activities and backward sensitivities are now matrices,
we no longer find a Kronecker product structure for the partial derivatives, i.e.\
$ %\begin{equation}
    \text{vec}\left( e^{(l)} a^{(l-1)\transpose} \right)
    \neq
    a^{(l-1)} \otimes e^{(l)} .
$ %\end{equation}
Instead, the $l^{\text{th}}$ block of the FIM in vectorised form now reads
\begin{align}
    \breve{F}^{(l)}(\theta)
    &=
    \mathbb{E}_{p_\theta}
    \left[
        \text{vec}\left(\frac{\partial \ln p_{\theta}}{\partial W^{(l)}}\right)
        \text{vec}\left(\frac{\partial \ln p_{\theta}}{\partial W^{(l)}}\right)^\transpose
    \right] \nonumber \\
    &=
    \mathbb{E}_{p_\theta}
    \left[
        \text{vec}\left( e^{(l)} a^{(l-1)\transpose} \right)
        \text{vec}\left( e^{(l)} a^{(l-1)\transpose} \right)^\transpose
    \right] \nonumber \\
    &=
    \mathbb{E}_{p_\theta}
    \left[
        \left(a^{(l-1)} \otimes e^{(l)}\right)
        \left(\text{vec}\left(I_A\right) \text{vec}\left(I_A\right)^\transpose\right)
        \left(a^{(l-1)} \otimes e^{(l)}\right)^\transpose
    \right] \ , \label{eq:FFN_fisher_block}
\end{align}
where $I_A$ is the $A \times A$ identity matrix.
In this result, we have used the general formula
$ %\begin{equation}
    \text{vec}\left( B C D \right)
    =
    (D^\transpose \otimes B) \text{vec}\left( C \right)
$~\cite{henderson1981vec}, %\end{equation}
which holds for any matrices $B$, $C$ and $D$ with compatible matrix products.
Comparing Eq.~\eqref{eq:fisher_block} to Eq.~\eqref{eq:FFN_fisher_block}, one
sees that the weight-sharing introduces an intermediate $A^2 \times A^2$ factor,
$\text{vec}\left(I_A\right) \text{vec}\left(I_A\right)^\transpose$,
which prevents us from recasting the FIM as the average of a Kronecker product.
More generally, as soon as a parameter of the NN contributes to more than one hidden node,
one must use the chain rule as given in Eq.~\eqref{eq:multidim_chain_rule}
and Eq.~\eqref{eq:fisher_block} does not hold any more.

Interestingly, this analysis suggests that the KFAC rationale remains valid for the $A=1$
case, i.e.\ for the particular case of one particle.
We expect that it will become less and less valid as $A$ increases.
This seems to support our empirical observation that QN-KFAC
performs better for small values of $A$.
The results obtained with QN-KFAC in Fig.~\ref{fig:KFAC_vs_QN-KFAC} look indeed more stable, 
and converge faster, for $A=2$ and $3$ than for $A\geq4$.

Following Eq.~\eqref{eq:optimisation_general_solution},
to perform the parameter update we have to solve the linear system
\begin{equation}\label{eq:MINRES_direction_update_linear_system}
    \left(\breve{F}(\theta_n) + \lambda_n I \right) \Delta^{M}_n = - \nabla E(\theta_n) \ .
\end{equation}
To do so, we employ the iterative linear solver MINRES with a maximum number of steps $N_\mathrm{MR}$, which we initialise at the direction obtained in KFAC. This allows us to systematically improve on the direction estimation. We dub the resulting optimiser QN-MR-KFAC. We note that when $N_{\mathrm{MR}} = 0$, QN-MR-KFAC reduces to QN-KFAC.
Let us also emphasise that, while the quasi-Hessian $H_Q(\theta)$ is used in the rescaling phase,
only the FIM is used in the evaluation of the direction of the update.
For clarity, we provide a schematic overview of the QN-MR-KFAC algorithm in Fig.~\ref{fig:masterFigureoptimisers}. More details about our implementation of MINRES
can be found in the Supplemental Material.

\begin{figure*}
\begin{center}
\includegraphics[width=0.49\linewidth, trim=0 100 0 100]{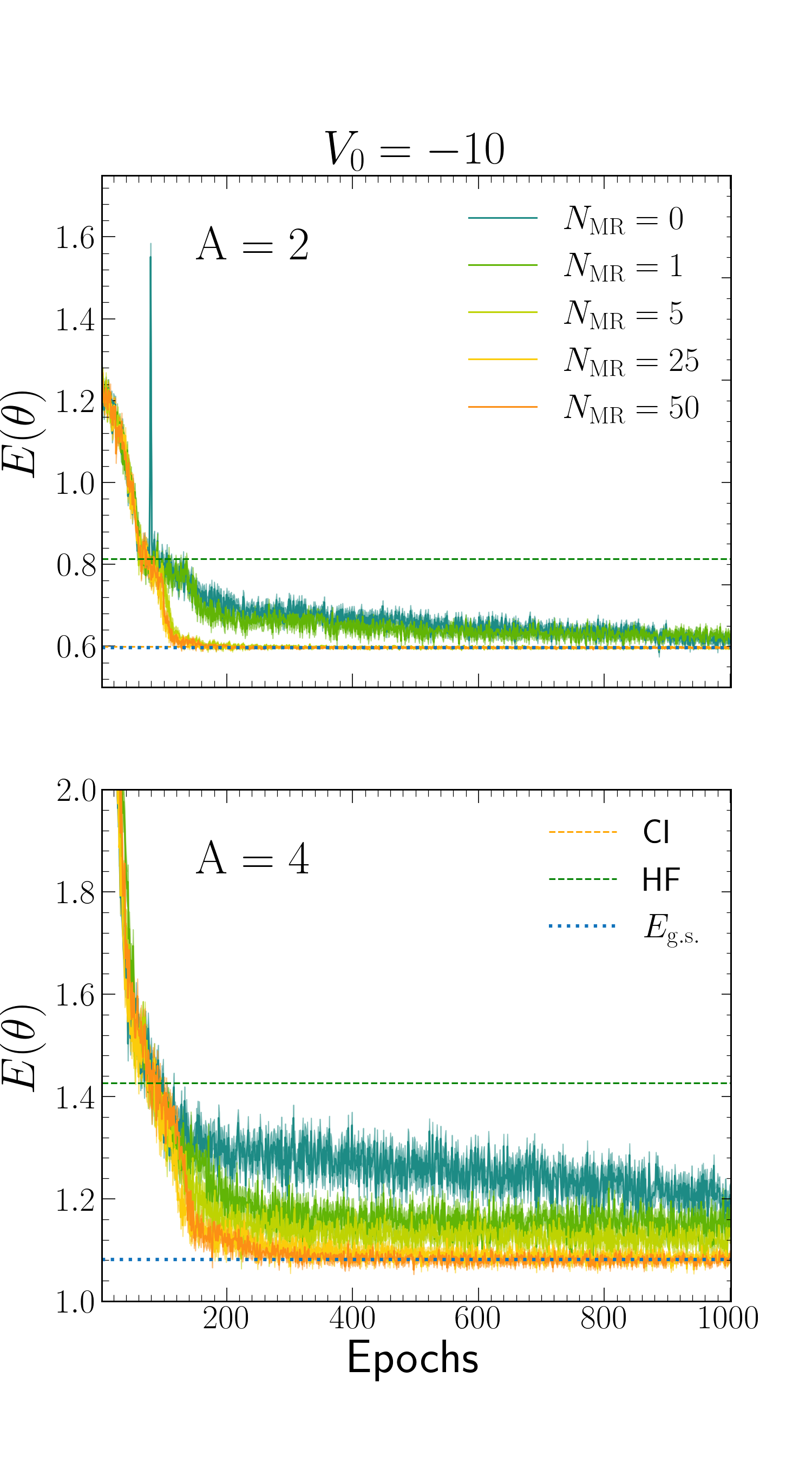}
\includegraphics[width=0.49\linewidth, trim=0 100 0 100]{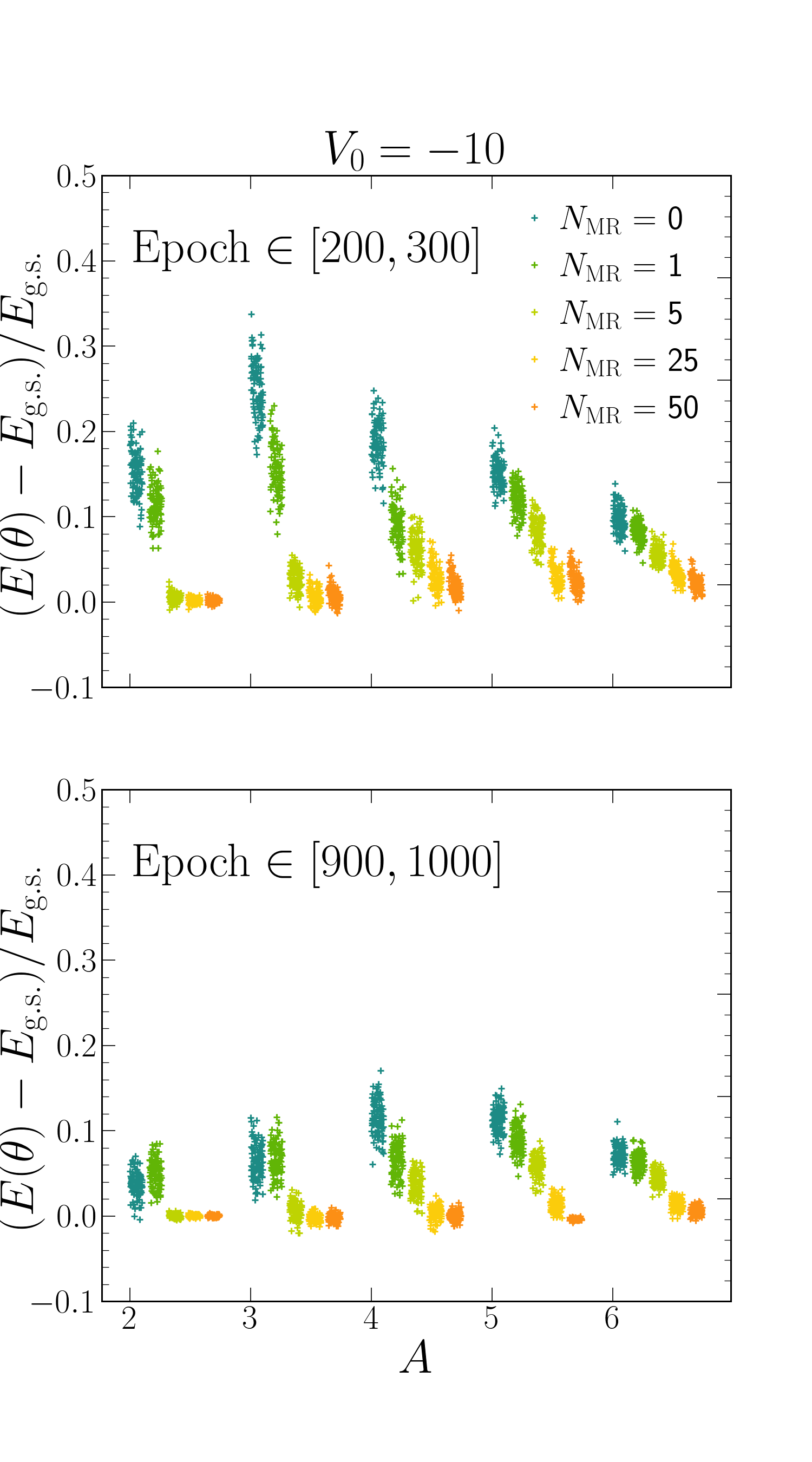}
\end{center}
	\caption{Evolution of the energy $E(\theta)$ as a function of the number of epochs for 
	 QN-MR-KFAC optimisations with $N_{\mathrm{MR}}=0,1,5,25,50$.
	 The top (bottom) left panels shows results for $A=2$ ($A=4$) for $V_0=-10$.
    HF and CI benchmarks are displayed for reference with green and orange dashed lines, respectively.
    Since we do not have access to converged CI result in all cases, we also display for reference
    results obtained in Ref.~\cite{keeble2023machine} with blue dotted lines and referred to as $E_{\mathrm{g.s.}}$.
    The top (bottom) right panels show results for $A=2,3,4,5$ and $6$ with epochs between $200$ and $300$ ($900$ and $1000$).
}
	\label{fig:QN-MINRES-KFAC_variable_iteration}
\end{figure*}

We now test the impact of using a refined directional update, $\Delta^M_n$, instead
of the bare KFAC estimate, $\Delta^\text{KFAC}_n$. 
We focus first on cases where QN-KFAC converges reasonably well
without violent oscillations. 
We show the energy as a function of the epoch number 
for $V_0=-10$ with $A=2$ (top left panel) and $A=4$ (bottom left panel) particles in
Fig.~\ref{fig:QN-MINRES-KFAC_variable_iteration}.
Specifically, we illustrate the impact of MINRES by exploring a different set of iterations,
$N_{\mathrm{MR}} \in \set{0, 1, 5, 25, 50}$.
To illustrate how the picture evolves as $A$ increases, we also show
snapshots of energies for epochs between $200$ and $300$ (top right panel)
and for epochs between $900$ and $1000$ (bottom right panel). 

The results clearly indicate that improving the direction
of the parameter update (or increasing $N_{\mathrm{MR}}$) has two key effects.
First, the oscillations in the energy estimate during the minimisation process are substantially reduced. 
Even for $A=2$ (top left panel), where the oscillations were relatively small, increasing $N_{\mathrm{MR}}$ leads 
to substantially reduced oscillations in the energy predictions. For $4$ particles (bottom left panel), QN-KFAC, which
corresponds to the $N_{\mathrm{MR}}=0$ line, has relatively large oscillations in energy, of order $0.1$ in HO units.
When $N_{\mathrm{MR}}$ increases to $5$, the oscillations are reduced by about a factor of $2$. A further increase
up to $N_{\mathrm{MR}}=50$ shows minor oscillations, to less than about $0.03$ in HO units.
Second, the convergence in the late stages of the optimisation is considerably faster. 
For $A=2$, QN-KFAC only converges to the CI result at around the $1000^\text{th}$ epoch. 
When the MINRES solver performs more than $25$ iterations per epoch, 
the optimisation process reaches the CI value in~$\approx 120$ iterations. 
In the $4$-particle case, the improvement is more dramatic. QN-KFAC plateaus, but does not converge,
at the $1000^\text{th}$ epoch. Increasing $N_{\mathrm{MR}}$, a more transparent convergence pattern appears
and, for $N_{\mathrm{MR}}=50$, about $400$ epochs are enough to reach the minimum. 
We stress that the characteristically slow approach to the minimum was
a major drawback of the QN-KFAC proposal. 

Overall, we observe that improving the quality of the direction of $\delta_n$ substantially impact
the performances. Comparing the $A=2$ and $A=4$ cases, we see that the number of MINRES iterations
necessary to reach optimal convergence is larger as $A$ increases.
This is further supported by the top right and top left panels of Fig.~4 which illustrate that, both in the early
and late stages of the optimisation, the number of MINRES iterations necessary to reach an optimal convergence increases with $A$.
This confirms our previous theoretical analysis expecting KFAC to be less and less well motivated as $A$ increases.
In practice, we find that setting $N_{\mathrm{MR}} = 50$ is sufficient for any case where a stable convergence is reached.

\begin{figure*}
\includegraphics[width=0.94\linewidth, trim=180 80 180 0]
{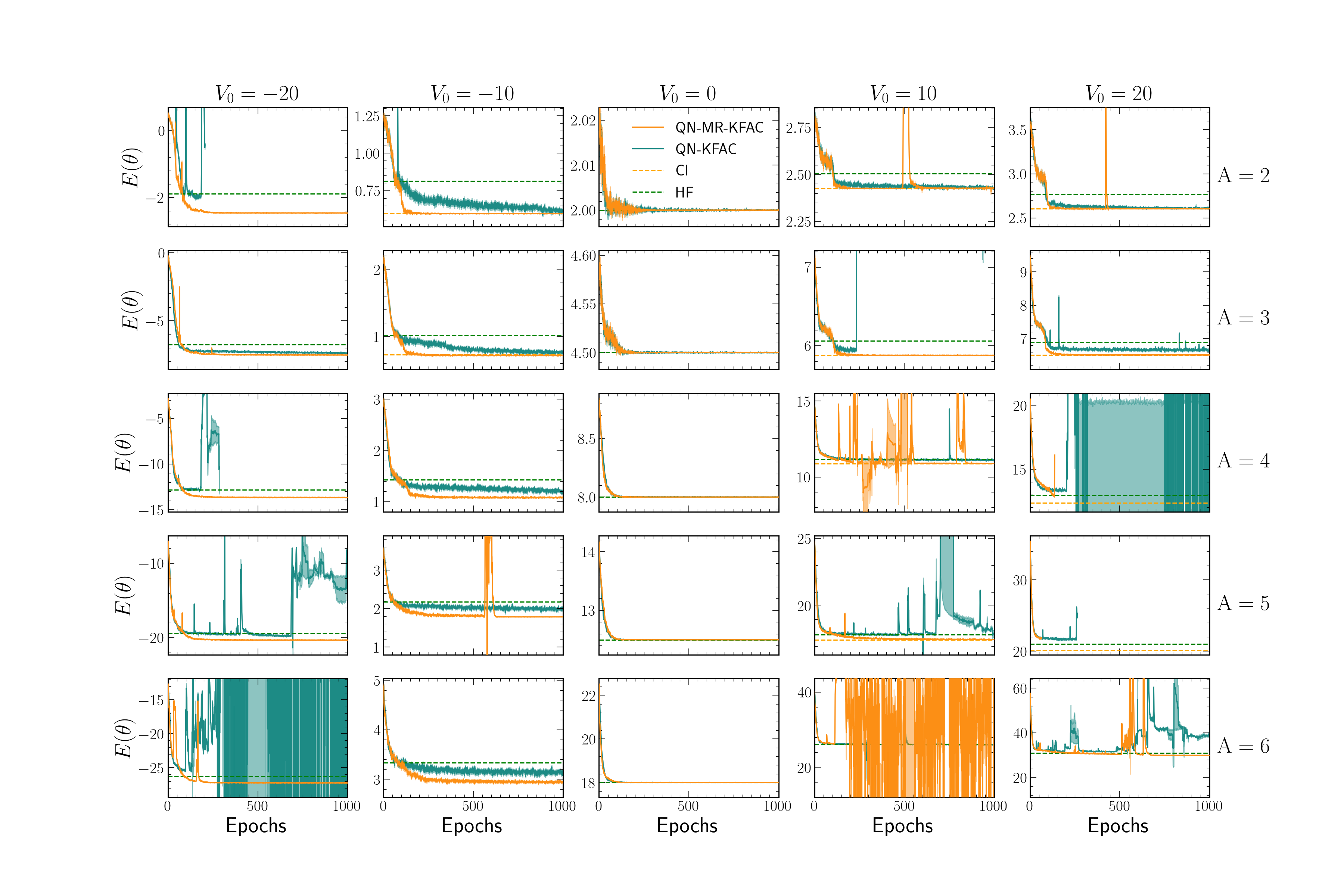}
	\caption{
	The same as Fig.~\ref{fig:KFAC_vs_QN-KFAC},
 but comparing the QN-KFAC optimiser (green solid line)
 to the QN-MR-KFAC optimiser with $N_\text{MR}=50$ (orange solid line). 
}
	\label{fig:QN-MINRES-KFAC_vs_QN-KFAC}
\end{figure*}

A more systematic analysis of the importance of the improvement in the direction of the updates
is provided in Fig.~\ref{fig:QN-MINRES-KFAC_vs_QN-KFAC}. Here, we show the same results
as in Fig.~\ref{fig:KFAC_vs_QN-KFAC}, but comparing QN-KFAC to QN-MR-KFAC with $N_\text{MR}=50$. 
The results confirm that QN-MR-KFAC greatly improves on both the oscillations of the energy values
and the late phase of convergence. In the vast majority of cases, QN-MR-KFAC performs better than
QN-KFAC in terms of accuracy and speed of convergence.
Concerning the stability of the optimisation process, the improvements are more marginal.
The attractive and non-interacting cases (columns to the left of, and including, the 
centre) only show some erratic peaks in energy in a few, selected cases. 
In contrast, the repulsive cases are still prone to important oscillations as soon as $A\geq4$. 

Finally, we notice that in several cases
(for example when $A=2$, $V_0=10$ and $A=5$, $V_0=-10$)
QN-MR-KFAC shows instabilities even \emph{after} having converged.
We interpret this  behaviour in terms of the dynamically adjusted damping.
Once convergence is reached, the trust region keeps increasing. One may eventually hit a remote spurious
point in the local quadratic model which has a value that is lower than the true ground-state energy and jump there.
The associated numerical instability takes a few epochs to self-adjust.
As stated before, this is one of the drawbacks of using a quasi-Hessian, $H_Q(\theta)$, which
is not positive semi-definite.
These pathological cases may in principle be cured by introducing a minimum value on the damping, $\lambda_n$.
More generally, most of the fluctuations encountered so far can be tackled by being more restrictive
on the trust regions. However, in doing so, we also restrict the size of the updates and slow down
the overall convergence of the optimiser. We also introduce the risk of an artificial convergence, in the sense that the energy
may converge only because the trust region has shrunk to a very narrow point. We leave further analysis of trust regions and their
dynamical adjustments for future work.

\section{Decisional gradient descent}
\label{sec:decisionalGradient}

\subsection{Failure of information geometry}
\label{sec:failure_info_geo}

Despite the multiple improvements to the original KFAC algorithm which we designed
to take into account the particularities of VMC and NQS,
our finest optimiser, QN-MR-KFAC, is still subject to violent fluctuations in the energy.
As discussed in Sec.~\ref{sec:direction_improvement}, this could be tackled by further
refining the trust regions. Doing so, however, may limit the original promise
of performance or require fine-tuning of the hyperparameters, as in past attempts to adapt
KFAC to VMC with NQS~\cite{pfau2020ab,wilson2021simulations}.
These attempts, including ours, rely on the use of the FIM, $F$, to estimate 
the direction of the updates $\Delta_n$.
The use of the FIM was originally motivated by its connection to the Hessian
of the loss function in a supervised learning setting.
We have already argued that standard VMC-NQS
does not fall in the category of supervised learning problems, and hence using the FIM (or
its corresponding approximations) is no longer justified.
The substantial improvement observed in Fig.~\ref{fig:KFAC_vs_QN-KFAC},
where the quasi-Hessian $H_Q$ is used in the rescaling phase of the optimiser
instead of the exact FIM $F$, further supports this idea.  
We hypothesise that, the failure of KFAC are not only due to the approximations
it makes to NGD, but also, at a more fundamental level, to the failure of the plain NGD
to provide a good optimiser for VMC when applied to the many-body systems we consider.

To test this hypothesis, we perform calculations with an optimiser that 
is similar to QN-MR-KFAC with $N_{\mathrm{MR}}=50$, except that the exact FIM, $F$, as 
opposed to the quasi-Hessian, $H_Q$, is used in the rescaling phase.
This is closer in spirit to Hessian-free optimisers~\cite{martens2012training}.
In addition, to avoid any contamination from KFAC, we replace the initial guess
in the MINRES solver by the previous update, rescaled by a prefactor $\zeta=0.95$, i.e.\
the starting point of the MINRES solver is now $\zeta \times \delta_{n-1}$ instead of $\Delta^{\text{KFAC}}_n$. 
This follows the default recommendation of Ref.~\cite{martens2012training} for Hessian-free optimisers.
All other hyperparameters are fixed to the same value used in our previous QN-MR-KFAC calculations.
Although we do not calculate exactly the natural gradient update, 
$\delta^{\text{NGD}}_n$ given in Eq.~\eqref{eq:NGD-update},
we deem this to be sufficiently close to the NGD approach. 
In consequence, we will refer to this optimiser as NGD.
A schematic overview of our implementation of NGD is provided in Fig.~\ref{fig:masterFigureoptimisers}.

\begin{figure*}
\includegraphics[width=0.94\linewidth, trim=180 80 180 0]{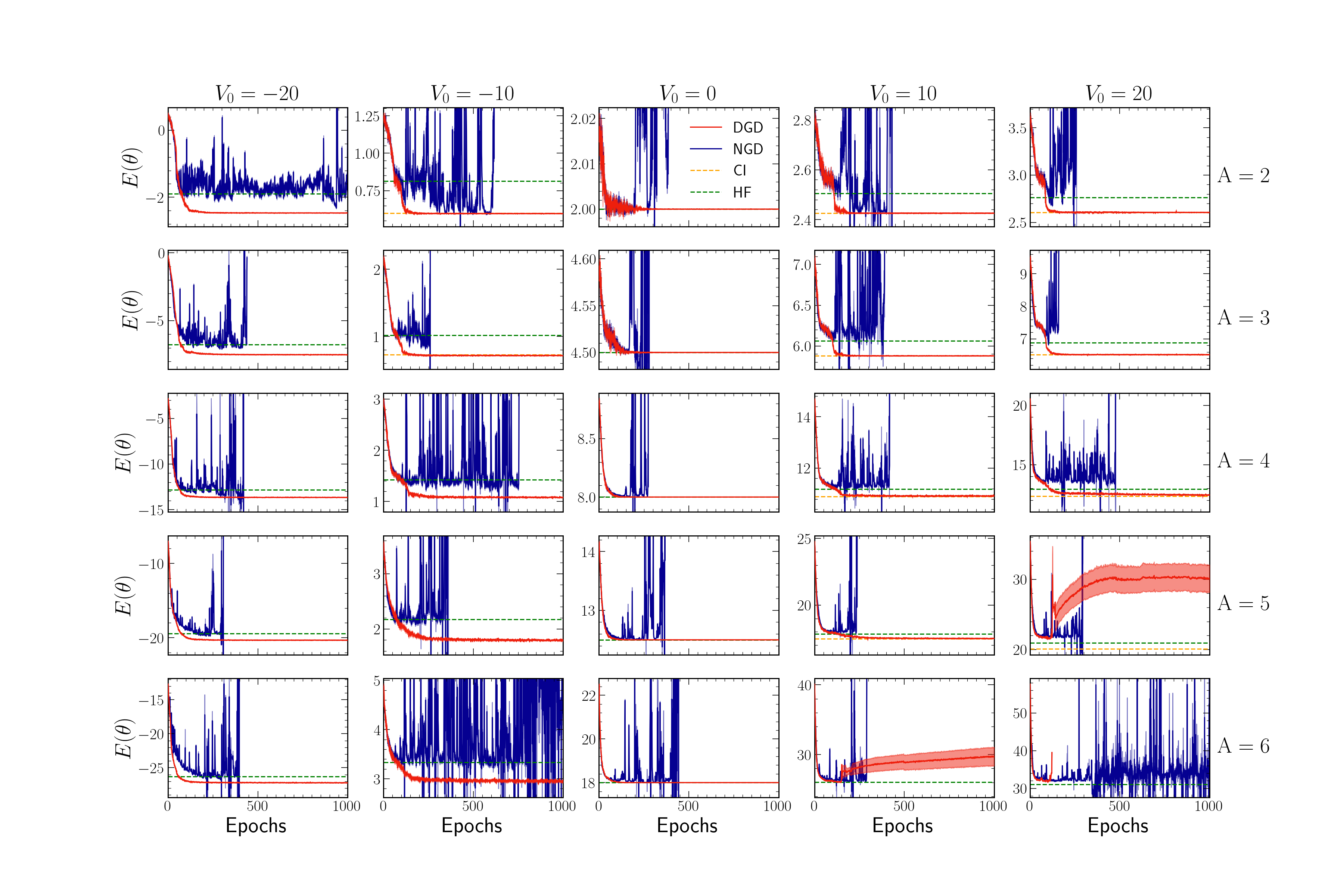}
	\caption{The same as Fig.~\ref{fig:KFAC_vs_QN-KFAC}, but comparing the NGD optimiser
    (dark blue solid line) to the DGD optimiser (red solid line) with $N_\text{MR}=50$. 
 }
	\label{fig:NGD_vs_DGD}
\end{figure*}

We present the results of this NGD optimiser in Fig.~\ref{fig:NGD_vs_DGD}.
The panels are organised in the same fashion as Fig.~\ref{fig:KFAC_vs_QN-KFAC}.
In all the cases considered, NGD fails to converge smoothly to a final value.
We find strong oscillations, akin to those observed
for KFAC in Fig.~\ref{fig:KFAC_vs_QN-KFAC}. In some cases,
particularly in the attractive regime (columns to the 
left of the centre), the optimiser seems to get stuck 
around the HF solution. Even then, violent oscillations
arise in the energy minimisation process. 
The generic failure of the NGD optimiser across different 
numbers of particles and interaction strengths further supports
our hypothesis that NGD does not fall in the category of second-order optimisers for VMC.
As discussed in Sec.~\ref{sec:NGD}, NGD can be seen as a simple gradient descent where the
definition of trust regions are based on information geometry,
i.e.,\ on the KL divergence~\eqref{eq:KLdiv}.
We interpret the failure of NGD as a consequence of the unsuitability of information 
geometry for VMC.

This raises the question: which geometry should one use for VMC?
Ideally, we would like to use a geometry leading to
(i) a local metric that is positive semi-definite, for stability purposes; and
(ii) a metric informed by the loss function that we are trying to minimise, here the energy $E(\theta)$.
While both conditions are satisfied
when considering information geometry for supervised 
learning problems, condition (ii) is violated in the case
of VMC simulations. In order to restore condition (ii),
we now introduce the concept of decision geometry.

\subsection{Decision geometry}

Our starting point consists in reformulating VMC in a game-theory framework.
In this paradigm, the standard information geometry
is generalised by the concept of
\emph{decision geometry}~\cite{topsoe1979information,grunwald2004game,dawid2007geometry,amari2016information}.

Similarly to how the KL divergence, Eq.~\eqref{eq:KLdiv}, is related to
the cross-entropy loss, Eq.~\eqref{eq:cross-entropy-loss},
one can define a divergence function associated to \emph{any} loss function $\mathcal{L}(\theta)$ of the type
\begin{equation}
    \mathcal{L}(\theta) \equiv -\mathbb{E}_{X \sim p_\theta}\left[\mathcal{S}(X, p_\theta)\right] \ ,
\end{equation}
so long as $\mathcal{S}(X, p_\theta)$ is a so-called 
\emph{proper scoring rule}~\cite{gneiting2007strictly}.
A scoring rule $\mathcal{S}(x, p)$ is any integrable function that takes as inputs a probability distribution, $p$,
and a sample, $x$, and returns a real valued quantity.
From any scoring rule, one defines the associated expected 
score $S(p,q)$ for any two probability distributions $p$ and $q$ by
\begin{equation}
    S(p,q) \equiv \mathbb{E}_{X \sim p}\left[\mathcal S(X, q)\right] \ .
\end{equation}
A scoring rule is said to be proper if and only if, for any two probability distributions $p$ and $q$,
\begin{equation}
    S(p,p) \leq S(p,q) \ .
\end{equation}
In this framework, several typical quantities of information theory can be generalised.
For example, the \emph{generalised entropy} and the \emph{generalised cross-entropy}
are simply $S(p,p)$ and $S(p,q)$, respectively.
Their associated divergence function, $D_S$, is defined by
\begin{equation}
    D_S(p, q) \equiv S(p,q) - S(p,p) \ .
\end{equation}
The new geometry, associated to $D_S$, is referred to as a
decision geometry~\cite{grunwald2004game,dawid2007geometry} because of its relation
with statistical decision problems~\cite{ferguson1967mathematical}.
Similarly to the KL divergence, $D_S$ defines a metric locally.
This is obtained by Taylor-expanding $D_S$ at second-order, namely
\begin{equation}
    D_{S}(p_{\theta} , \ p_{\theta+\delta})
    = \frac{1}{2} \delta^\transpose G_S(\theta) \delta  + O\left(\delta^3\right)\ ,
\end{equation}
where $G_S(\theta)$ is the decision metric associated to $S$.
Let us emphasise that, by construction, $G_S(\theta)$ is symmetric positive semi-definite.

This symmetric and positive definite decision metric 
provides a natural starting point for optimisation schemes that
are not tied to the FIM. 
In this context, we define a novel optimiser, the \emph{decisional gradient descent} (DGD)
which consists simply in performing a gradient descent 
with trust regions
defined by the norm associated to the decision metric, 
$ %\begin{equation}
    T^{\text{DGD}}_n = \set{\delta : \delta^\transpose G_S(\theta_n) \delta \leq r^2} \ .
$ %\end{equation}
Here, $r > 0$ denotes the maximum size on the update $\delta_n$.
Just like for NGD, it is more practical to work with an equivalent 
unconstrained regularised problem. This leads to the definition
of an update,
\begin{equation}
    \delta^{\text{DGD-reg}}_n = - \alpha \ G_S^{-1}(\theta_n) \nabla \mathcal{L}(\theta_n) \ ,
\end{equation}
where $\alpha$ is again a learning rate that can be related to the size of $T^{\text{DGD}}_n$. 

If we choose the scoring rule $S^{\text{sup}}(X,p_\theta) \equiv -\ln{p_\theta(X)}$,
the decision geometry framework is such that one recovers exactly 
the standard cross-entropy, Eq.~\eqref{eq:cross-entropy-loss}, and
the KL divergence, Eq.~\eqref{eq:KLdiv}. In other words, DGD reduces to NGD.
To go beyond NGD, we move away from the standard logarithmic scoring rule and
choose instead an alternative one based on the local energy, namely
\begin{align}
    \mathcal{S}_{\text{VMC}}(X,p_\theta)
    &\equiv - E_{L,\theta}(X),  \label{eq:vmc_score}
\end{align}
where $p_\theta = \abs{\Psi_\theta}^2$ is the Born probability associated to the many-body wave function.
We refer to this scoring rule as the \emph{VMC scoring rule}.
The local energy in our toy model is explicitly given in Eq.~\eqref{eq:local_energy}.
One can show that, for such local energy,
the inequality
\begin{equation} \label{eq:elocal_proper}
    \mathbb{E}_{X \sim p_{\theta}}
    \left[  -E_{L,\theta}(X) \right]
    \leq
    \mathbb{E}_{X \sim p_{\theta}}
    \left[  -E_{L,\theta'}(X) \right] \ 
\end{equation}
holds for probability distributions $p_\theta$ and $p_\theta'$ in $\mathcal{P}$~\cite{ehm2012local}.
The class $\mathcal{P}$ contains all probability distributions that are smooth,
that vanishes fast enough to infinity and that does not
cancel out on any finite $X$. 
In other words, $\mathcal{S}_{\text{VMC}}$ is a proper scoring rule with respect to $\mathcal{P}$.
Generalisations of $\mathcal{S}_{\text{VMC}}$ to make it proper with respect
to any physical probability distributions is left for future work.
Following the analogy with information geometry, the total energy 
$E(\theta)$, computed as the expectation value of the local energy, 
is analogous to a generalised entropy
\begin{equation}
    E(\theta) = - S_{\text{VMC}}(p_\theta, p_\theta) \ ,
\end{equation}
with associated divergence
\begin{align}\label{eq:vmc_divergence}
    D_{\text{VMC}}(p_\theta, p_{\theta'})
    &=
    \mathbb{E}_{X \sim p_{\theta}}
    \left[  E_{L,\theta}(X) - E_{L,\theta'}(X)\right] \ . 
\end{align}
For our toy model, employing continuous variables in a one-dimensional setting and a local interaction term, 
Eq.\eqref{eq:local_energy}, this difference reduces to the
expectation value
\begin{align}
    D_{\text{VMC}}(p_\theta, p_{\theta'})=
    \frac{1}{8} \sum_{k=1}^A \mathbb{E}_{X \sim p_{\theta}}
    \left[
        \left(
            \partial_{x_k} \ln{p_{\theta}(X)} - \partial_{x_k} \ln{p_{\theta'}(X)}    
        \right)^2
    \right] \ . \label{eq:VMCdiv}
\end{align}
This divergence arises entirely from the kinetic term of the
Hamiltonian, which is the reason derivatives appear. 
We note that this divergence can be written in terms of first-, rather than second-order derivatives, because we assume regularity at the boundaries.
For our neural network, it can be analytically proven that the boundary terms appearing in the derivation of the divergence are zero,
and therefore Eq.~\eqref{eq:VMCdiv} holds.
However, this need not be the case for arbitrary NN architectures or distributions.
A more detailed derivation of this expression can be found in the Supplemental Material.

As can be seen from Eq.~\eqref{eq:vmc_divergence}, the decision geometry obtained from the VMC scoring rule defined in Eq.~\eqref{eq:vmc_score}
provides a connection between the geometry and the loss function we aim to minimise, $E(\theta)$.
Indeed, the divergence we obtain can be seen as an approximation of the difference of energy $E(\theta) - E(\theta')$,
where we would have neglected re-weighting the samples.
The metric, $G_{\text{VMC}}(\theta)$, obtained from
Taylor-expanding the divergence thus fulfils the two main theoretical requirements
we discussed in Sec.~\ref{sec:failure_info_geo}, namely
(i) the metric is symmetric positive definite and 
(ii) the metric is informed by the loss function $E(\theta)$ we aim to minimise.
The metric reads explicitly
\begin{equation}
    G_{\text{VMC}}(\theta)_{\theta_i \theta_j}
    \equiv
    \frac{1}{4}
    \sum_{k=1}^A
    \mathbb{E}_{X \sim p_\theta}
    \left[
        \partial_{\theta_i}\partial_{x_k} \ln p_{\theta}(X) \ 
        \partial_{\theta_j}\partial_{x_k} \ln p_{\theta}(X)
    \right] \ , \label{eq:VMCmetric}
\end{equation}
and we will use it later on in simulations. We direct the interested reader to the Supplemental Material for a proof for this expression. A peculiarity of this equation
is the appearance of mixed derivatives with respect to both parameters 
and inputs of the probability distribution. We will provide more details on the DGD optimiser in a forthcoming publication~\cite{Keeble2024}.

Interestingly, using the natural scoring rule of Eq.~\eqref{eq:vmc_score} 
for VMC leads to a geometry that is equivalent to the one introduced by 
Hyvärinen from very different initial considerations~\cite{hyvarinen2005,hyvarinen2007,amari2016information}.
In Ref.~\cite{hyvarinen2005}, the main concern was to bypass the difficulty of sampling according to a non-normalised probability distribution. 
The so-called Hyvärinen score has the following form
\begin{equation}
    \mathcal{S} _{\text{H}}(X, p_\theta)
    \equiv
    \sum_{k=1}^A \left[ 
     \partial^{2}_{x_k} \ln{p_\theta} (X) + 
     \frac{1}{2} \left( \partial_{x_k} \ln{p_\theta}(X) \right)^{2} \right] \ ,
\end{equation}
and it is essentially the kinetic part of the local energy, Eq.~\eqref{eq:local_energy}. 
Our VMC score is related to the Hyvärinen score through the formula
$ %\begin{equation}
    \mathcal{S}_{\text{VMC}}(X, p_\theta) = 
    \frac{1}{4} \mathcal{S}_{\text{H}}(X, p_\theta) - V(X) \ .
$ %\end{equation}
As a result, the associated divergence function and metric are proportional to each other,
i.e.\ their decision geometry are identical up to a scaling factor.
In the language of game theory, the Hyvärinen and VMC scoring rules are said to be
\emph{equivalent}~\cite{dawid2007geometry,gneiting2007strictly}.

\subsection{Application to VMC}

We now proceed to test a DGD-based optimisation strategy on VMC calculations with NQS.
In this case, one uses the VMC scoring rule, Eq.~\eqref{eq:vmc_score}, which defines the metric $G_{\text{VMC}}$.
Moreover, the loss function is the energy of the system, so the update in the parameters immediately reads
\begin{equation}
    \delta^{\text{DGD-reg}}_n = - \alpha \ G_{\text{VMC}}^{-1}(\theta_n) \nabla E(\theta_n) \ .
\end{equation}
In order to provide a fair comparison with NGD and the associated
approximated optimisers (KFAC, QN-KFAC or QN-MR-KFAC), we have set the 
learning rate $\alpha = 1$.
This common choice is critical to ensure that any success of DGD 
is well motivated and that DGD can be interpreted as a second-order optimiser.
In this case, one solves the sequence of quadratic sub-problems
\begin{equation}\label{eq:DGD_subproblem}
    M^{\text{DGD}}_n(\delta) =
    \frac{1}{2} \delta^\transpose G_{\text{VMC}}(\theta_n) \delta
    + \nabla E(\theta_n)^\transpose \delta + E(\theta_n) \ .
\end{equation}
As a first implementation of DGD, we use the very same code employed in our
NGD optimiser. The difference is that we replace 
the FIM $F(\theta_n)$ in the NGD, by the decision metric,
$G_{\text{VMC}}(\theta_n)$.
We provide a schematic overview
of the implementation of DGD in Fig.~\ref{fig:masterFigureoptimisers} and remind the reader that more information on hyperparameters can be found in the Supplemental Material.

Results for the NGD and DGD optimisers are shown hand in hand in Fig.~\ref{fig:NGD_vs_DGD}. 
Clearly, the stability of the optimisation process is dramatically
improved when employing DGD. A well-defined initial minimisation occurs
for the first $100$ epochs in all cases. Importantly, the energy minimum is 
usually reached within this first $100$ epochs. 
No strong fluctuations are observed
in the energy either close or away from the minimum. Out of the $25$ cases
presented in Fig.~\ref{fig:NGD_vs_DGD}, only three extreme cases (i.e. about
$12 \%$ of the minimisations) fail to converge. Explicitly, these cases 
correspond to 
$A=5$ particles with $V_0=20$ and $A=6$ with $V_0=10$ and with $V_0=20$. 
In the first two cases, the lack of convergence is seen in terms of an
energy drift at large numbers of epochs, which happens after a short plateau close to the physical minimum. In the latter case, a numerical instability
precludes any minimisation past $100$ epochs. 
Let us stress that all the simulations shown in Fig.~\ref{fig:NGD_vs_DGD} 
employed the exact same set of hyperparameters (see Supplemental Material). 
In this context, it does not seem unreasonable to expect that further 
refinements of the DGD optimisation algorithm may lead to a 
systematic convergence to the ground-state energy.

\begin{figure*}
\includegraphics[width=0.94\linewidth, trim=180 80 180 0]{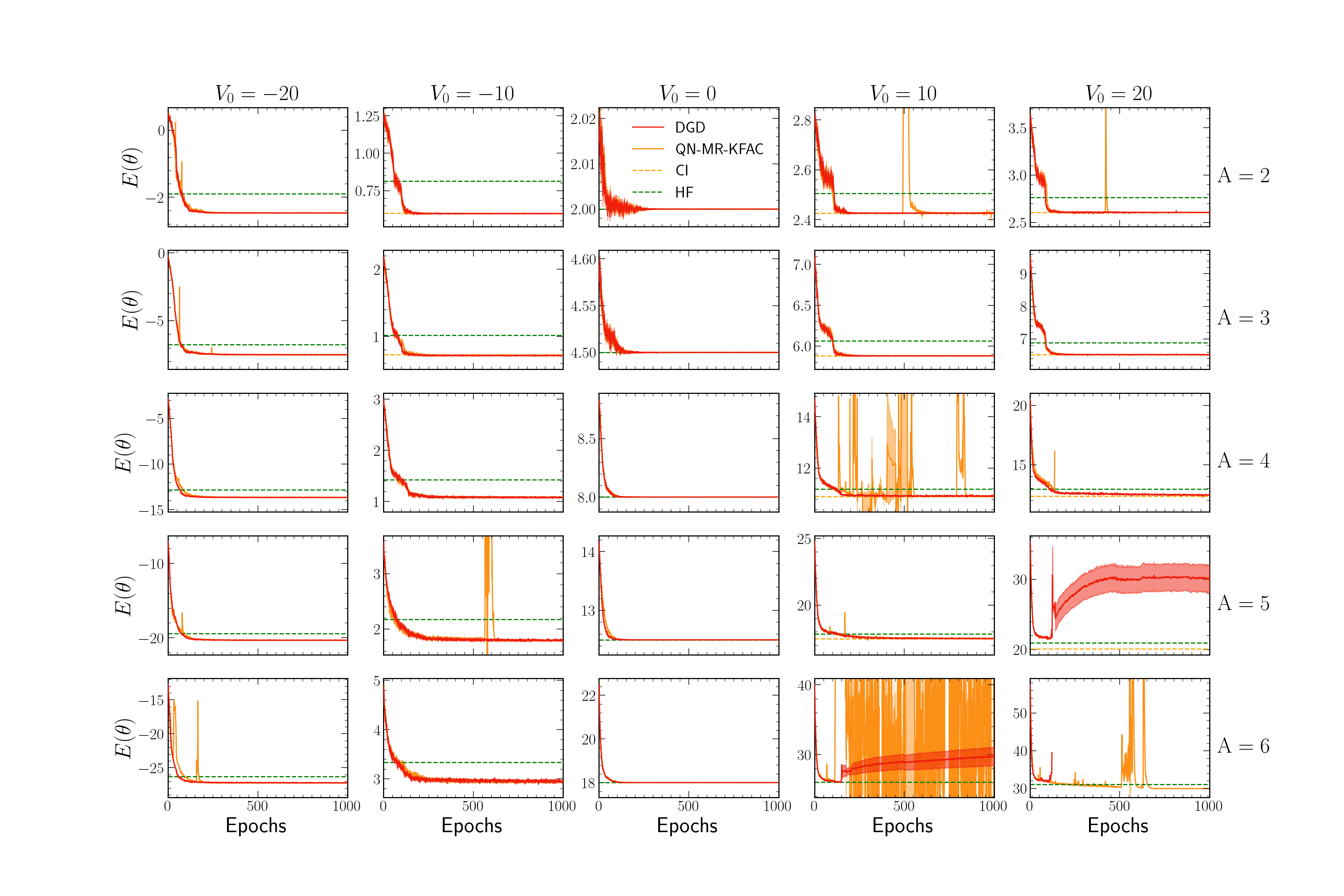}
	\caption{
 The same as Fig.~\ref{fig:KFAC_vs_QN-KFAC}, but comparing the DGD optimiser
    (red line) to the QN-MR-KFAC optimiser (orange solid line) with $N_\text{MR}=50$. 
 }
	\label{fig:QN-MR-KFAC_vs_DGD}
\end{figure*}

We provide further insight on the performance of the DGD optimiser in 
Fig.~\ref{fig:QN-MR-KFAC_vs_DGD}, where we compare it to our best 
NGD-based optimiser, QN-MR-KFAC.
Overall, we observe that the DGD optimiser generally performs as well as, if not better than, 
QN-MR-KFAC in terms of accuracy and speed of convergence. In the attractive 
regime, corresponding to the panels left and including the 
central column, we find 
that the two optimisers perform almost equally well. 
DGD, however,  improves
on the stability of the minimisation, avoiding the jumps that 
QN-MR-KFAC shows during the minimisation process. Similar conclusions
hold for the repulsive regime (panels right to the central column), 
although some cases with $A\geq5$ are challenging for 
both optimisers.

Compared to NGD, we interpret the overall improvement of DGD in terms of the direct
link that is established between the local model and the loss
function. 
Compared to QN-MR-KFAC, 
the fact that we keep a positive semi-definite metric, unlike the quasi-Hessian of Eq.~\eqref{eq:quasi-hessian}, is also an advantage
that avoids the instabilities discussed 
in Sec.~\ref{sec:direction_improvement}.

As an additional example that showcases the optimisation power of DGD, 
we now proceed to compare DGD results with Adam~\cite{kingma2014adam}.
We have relied on Adam in the NQS applications of 
Ref.~\cite{keeble2023machine}, where we employ
the default hyperparameters and a learning rate of $10^{-4}$. 
In that case, we train the network to minimise the energy for $10^5$ epochs. 
This relatively long training process is required to guarantee the
convergence of the energy, which is particularly slow in the last 
part of the training, when approaching the energy minimum. 

\begin{figure*}[t]
\begin{center}
\includegraphics[width=0.85\linewidth, trim=0 30 0 0]
{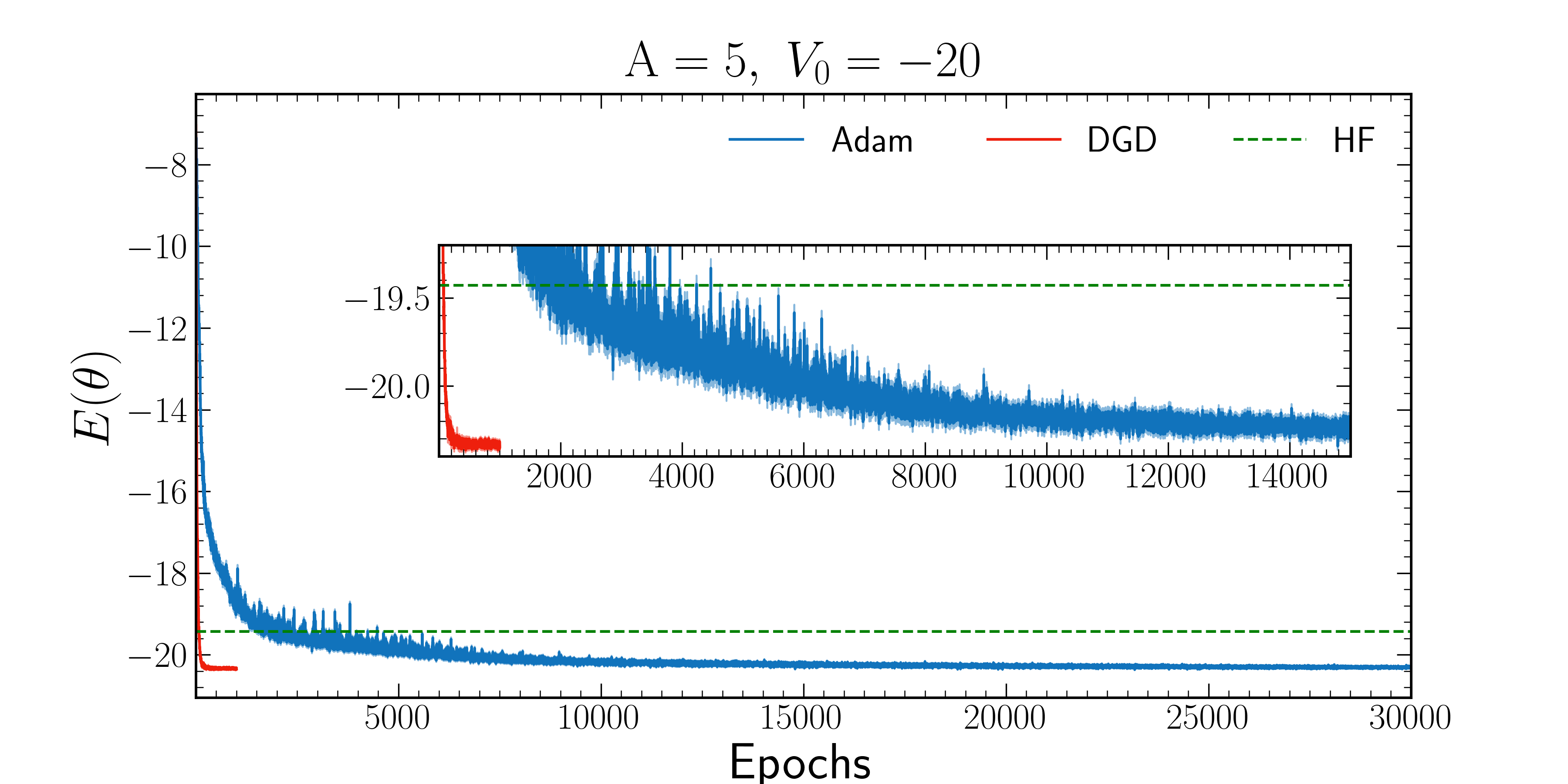}
\end{center}
	\caption{Evolution of the energy $E(\theta)$ as a function of the number of epochs for Adam
    (solid blue line) and for DGD (solid red line) with $N_{\mathrm{MR}}=50$, in
    the case of $A=5$ and $V_0=-20$.
    }
	\label{fig:accuracy_DGD_vs_Adam}
\end{figure*}

In Fig.~\ref{fig:accuracy_DGD_vs_Adam}, we display illustrative results 
for $A=5$ and $V_0=-20$ for Adam and DGD.
The results show
that the number of epochs required to reach a similar level of accuracy 
in the energy minimum is dramatically reduced when using DGD. 
Such speed-up is even more pronounced
when we focus on the correlation stage i.e.\ 
when comparing the number of epochs to converge, starting from the epoch
where the HF energy has been reached. 
Interestingly, the fluctuations on the energy value between epochs 
are also substantially reduced. DGD achieves
oscillations that are well within $0.1$ in HO units 
within a few hundred epochs.
In contrast, Adam only reaches a similar level of accuracy well past $20000$ epochs.

We note that we have performed more extensive comparisons between 
DGD and Adam
and we show these in the Supplemental Material. 
These extensive comparisons indicate that, while
the number of epochs that Adam requires to converge increases with $A$,
the performance of DGD stays relatively constant.
This should be helpful for tackling larger system in future NQS simulations.
Overall, compared to Adam, DGD shows a strong adaptability to the optimisation problem
and a particularly high performance in the late stage of convergence.
Those two empirical features are typical of second-order optimisers~\cite{martens2020new}.

Finally, let us also comment on the wall-time associated to the different methods.
DGD brings in a substantial acceleration in terms of wall-time, thanks to the reduction in the number of epochs. 
A timing profile of our NQS implementation~\cite{Azzam2023} indicates that the dominant time factor is the Markov Chain Monte Carlo (MCMC) propagation. 
Our MCMC is particularly demanding, because
 we have run  simulations with a large number of
 MH steps in order to avoid autocorrelation issues.
It is likely that 
the wall-time per epoch can be reduced further 
by setting a more reasonable number of MH iterations per epoch. 
This would affect the performance of both Adam and DGD. 
Further quantification of the numerical complexity of the DGD optimiser 
are left for future work.

\section{Conclusion and outlook}

NQS are a promising new tool in quantum many-body physics. In nuclear
physics, these methods hold the promise to become a competitive \emph{ab initio}
approach, tackling nuclei from first principles. In particular, given their VMC underpinning, these methods 
may reach relatively large particle numbers with a commensurate 
computational cost. The extension of NQS techniques to 
heavier systems requires technical developments for the energy
optimisation process. Here, we have focused on improving existing 
state-of-the-art optimisation schemes for NQS. 

We have compared several 
optimisation schemes  on a one-to-one basis.
We employ the setting and toy model of Ref.~\cite{keeble2023machine} as a
test bed, which has a rich phenomenology in spite of its simplicity.
We hope that the conclusions we draw can be extrapolated to
other systems. We mainly focus on the KFAC algorithm, that has been
 extensively employed in past NQS simulations~\cite{pfau2020ab}. Our numerical analysis 
 indicates that KFAC cannot be employed safely in a general VMC-NQS setting. We find
 a series of large oscillations in the minimisation process for a wide variety of physical properties. 
 KFAC was originally designed for tackling supervised learning problems
 with feed-forward NNs. Instead, VMC falls into the category of reinforcement learning problems
 and NQS architectures are rarely simple feed-forward networks.
 New and original approaches are necessary in order to recover
 the performance of second-order optimisers within a general VMC-NQS setting.

 In an effort to remedy the situation, we first propose two variants of 
 KFAC that try to bridge the gap towards VMC-NQS. 
 First, to account for the fact that VMC is not a supervised learning problem,
 we replace the exact FIM used in the rescaling phase by an approximation of the Hessian
 of the energy, which can be computed easily but is not positive-definite.
 The resulting approach, dubbed QN-KFAC, is more stable than the original KFAC
 implementation, but its convergence is still erratic in half the cases. 
 Second, to adapt KFAC to a general NQS, we propose an improvement in 
 the direction estimation of the parameter updates using the MINRES iterative linear solver.
 This allows us to mitigate the breaking of the Kronecker factorisation due to
 the permutation-equivariant structure of our NQS.
 The resulting optimisation scheme, QN-MR-KFAC, is much more stable and accurate than any of the previous optimisers.
 In spite of the improvements, this optimiser still shows instabilities and spikes in energy, even close to the minimum. 

Our third and last attempt to improve the optimisation of NQS is a 
radical departure from  KFAC. Instead of employing information geometry 
to inform the learning process of our model, we take a decisional 
approach based on game theory developments. This allows us to design
an optimisation scheme that is tailored to our VMC calculations.
The associated decision geometry defines an alternative positive semi-definite metric
which restores the link between geometry and loss function.
This so-called DGD method has a similar computational
cost to others, but provides a much faster, and much more stable, 
minimisation pattern. A relatively simple implementation of DGD shows already
a fast and stable convergence in more than $85\%$ of the cases. 
Because our toy model is relatively generic but incorporates a wide phenomenology, we expect similar improvements in other quantum many-body systems.
In particular, we hope that our optimisation strategies will be useful in developing NQS further for nuclear physics applications.  

The overall stability of DGD indicates that it has superior capabilities when it comes to VMC as compared to KFAC
or any of the resulting approaches we derived. 
Many improvements on this first
implementation of DGD can already be anticipated. To mention just a few, one could improve
the stability and speed of convergence by implementing a better damping scheme, following Ref.~\cite{martens2012training}.
The overall wall-time could also be reduced by replacing the iterative linear solver, MINRES, with another one
minimising directly the quadratic sub-problem in Eq.~\eqref{eq:DGD_subproblem}.
Recently, some groups have also proposed alternative approximations to KFAC that aim at reducing the cost of
solving linear systems at each epoch~\cite{chen2023efficient,rende2023simple,goldshlager2024kaczmarzinspired}.
Adapting those techniques to DGD could lead to important reductions of the wall-time per epoch. 

We note also that other attempts at designing better metrics than the FIM have recently appeared ~\cite{sasaki2024quantum}.
In this context, decision geometry provides a framework of choice due its high degree of flexibility.
In particular, it is intriguing that the potential energy, a crucial characterisation of the optimisation problem,
cancels out in the generation of the divergence in Eq.~\eqref{eq:VMCdiv}. Building new divergence functions incorporating information
on the potential energy could be an interesting avenue to further refine the geometry to the particular optimisation problem.
We are currently working on exploiting this flexibility to
study spin lattice systems~\cite{Keeble2024}.
Finally, beyond considerations on VMC and NQS, the versatility of decision geometry suggests
that it could be a promising avenue to design fast and robust optimisation strategies
for a broad class of machine learning problems.

\section*{Conflict of Interest Statement}

The authors declare that the research was conducted in the absence of any commercial or financial relationships that could be construed as a potential conflict of interest.

\funding{
This work is supported by STFC, through Grants Nos ST/L005743/1 and ST/P005314/1; 
by grant PID2020-118758GB-I00 funded by MCIN/AEI/10.13039/501100011033; 
by the ``Ram\'on y Cajal" grant RYC2018-026072 funded by MCIN/AEI /10.13039/501100011033 and FSE “El FSE invierte en tu futuro”; 
by the ``Consolidaci\'on Investigadora" Grant CNS2022-135529 
funded by MCIN/AEI/ 10.13039/501100011033 and by the “European Union NextGenerationEU/PRTR”;
by the ``Unit of Excellence Mar\'ia de Maeztu 2020-2023" award to the Institute of Cosmos Sciences, Grant CEX2019-000918-M funded by MCIN/AEI/10.13039/501100011033.
TRIUMF receives federal funding via a contribution agreement with the National Research Council of Canada. 
This work has been financially
supported by the Ministry of Economic Affairs and Digital
Transformation of the Spanish Government through the
QUANTUM ENIA project call - Quantum Spain project,
and by the European Union through the Recovery, Transformation
and Resilience Plan - NextGenerationEU within the
framework of the Digital Spain 2026 Agenda.
The authors gratefully acknowledge the computer resources at Artemisa, funded by the European Union ERDF
and Comunitat Valenciana as well as the technical support provided by the Instituto de Fisica Corpuscular, IFIC (CSIC-UV).
}

\section*{Supplemental material}

We provide a Supplemental Material file.
 
\section*{Data, code and materials}

The source code used for this study can be found in the GitHub repository {\url{https://github.com/jwtkeeble/second-order-optimization-NQS}}.
% The datasets [GENERATED/ANALYZED] for this study can be found in the [NAME OF REPOSITORY] [LINK].

\section*{Acknowledgements}

The authors would like to thank Richard Zou and the rest of the functorch team for their support and fruitful discussion in using the \texttt{torch.func} namespace.
% , which helped this research be computationally feasible. 

\bibliographystyle{RS} 
\bibliography{bibliography}

%%% Make sure to upload the bib file along with the tex file and PDF
%%% Please see the test.bib file for some examples of references
\end{document}

\jname{rsta}
\Journal{Phil. Trans. R. Soc. A}

%%%% Article title to be placed here
\title{Supplemental Material:
"Second-order optimisation strategies for neural network quantum states"}

\author{%%%% Author details
M.~Drissi$^{1}$,
J.~W.~T.~Keeble$^{2}$,
J.~Rozalén Sarmiento$^{3,4}$ and 
A.~Rios$^{2,3,4}$}

%%%%%%%%% Insert author address here
\address{
$^{1}$TRIUMF, Vancouver, V6T 2A3, British Columbia, Canada
\\
$^{2}$ Department of Physics, University of Surrey, Guildford GU2 7XH, United Kingdom
\\
$^{3}$ Departament de F\'isica Qu\`antica i Astrof\'isica, 
 Universitat de Barcelona (UB), 
 c. Mart\'i i Franqu\`es 1, E08028 Barcelona, Spain
\\
$^{4}$ Institut de Ci\`encies del Cosmos (ICCUB),
Universitat de Barcelona (UB), 
Barcelona, Spain
}

%%%% Subject entries to be placed here %%%%
\subject{quantum physics, nuclear physics, artificial
intelligence, game theory}

%%%% Keyword entries to be placed here %%%%
\keywords{neural network quantum states, Variational Monte Carlo, game theory, decision geometry, optimisation}

%%%% Insert corresponding author and its email address}
\corres{M.~Drissi\\
\email{mdrissi@triumf.ca}}

\begin{fmtext}
\section{MINRES}{\label{app:minres}}
In the main text we argue that the original arguments that motivate the Kronecker factorisation in KFAC do not hold in NQS. To compensate for this incompatibility, we introduce an additional step in QN-KFAC. Our goal is to improve 
on the direction of the update $\Delta^\text{KFAC}_n$
starting from the one obtained from KFAC. To do so, we use an iterative solver for linear equations. We employ MINRES~\cite{paige1975solution}, an algorithm that solves linear systems,
\begin{equation}\label{eq:linear_system}
    M x = y,
\end{equation}
for sparse, symmetric matrices, $M$, and output vectors, $y$.
The idea of MINRES is to combine a Lanczos algorithm with an LQ factorisation
in order to minimise the squared residual
\begin{equation}\label{eq:residual_minres}
    r(x)^2 \equiv (Mx - y)^\transpose (Mx - y) . 
\end{equation}
The Lanczos part approximately tridiagonalise $M$ as
\begin{equation}
    T_k = V_k^\transpose  M  V_k \ ,
\end{equation}
where $V_k$ denotes here the matrix of Lanczos vectors after $k$ iterations
and $T_k$ denotes the associated tridiagonal matrix containing the Lanczos coefficients.
Both matrices are of size $k \times k$. Given the fact that we enforce $k \leq N_{\mathrm{MR}}$
we are guaranteed that $T_k$ is at most of the much smaller size
$N_{\mathrm{MR}} \times N_{\mathrm{MR}}$. The rest of the MINRES algorithm
\end{fmtext}

\maketitle
\noindent
consists then in applying a LQ factorisation on $T_{k}$ and using the obtained
factors to build an approximate solution of the linear system Eq.~\eqref{eq:linear_system}.
Since the matrices $T_k$ are of small size this part of the algorithm is usually
of a negligible numerical cost as long as $N_{\mathrm{MR}}$ can be kept relatively small.
For more details on MINRES we refer to Ref.~\cite{paige1975solution}.

In our case, we use MINRES to find an approximate solution to the system
\begin{equation}
\label{eq:MINRES_direction_update_linear_system}
    \left(\breve{F}(\theta_n) + \lambda_n I \right) \Delta^{M}_n = - \nabla E(\theta_n) \, ,
\end{equation}
using $\Delta^\text{KFAC}_n$ as an initial guess. We recall that  
$\breve{F}$ is the block-diagonal approximation of the FIM
as defined in the main text.
% For simplicity, we choose to reuse $\lambda_n$ to regularise the directional update
% improvement, instead of $\gamma_n$.
Since we use the block-diagonal FIM \emph{without} the KFAC in Eq.~\eqref{eq:MINRES_direction_update_linear_system}
we consider it to be reasonable to choose the same trust region used for the rescaling (which uses the exact FIM).
Following Ref.~\cite{martens2012training}, we also
use the preconditioner $\left( \text{diag}(\breve{F}(\theta_n)) + \kappa I \right)^\xi$ with
$\kappa=10^{-2}$ and $\xi=0.75$ to accelerate the convergence of the linear solver.
In addition, MINRES is stopped as soon as we achieve a relative residual smaller than $10^{-8}$
or when the number of iteration reaches a maximum number, $N_{\mathrm{MR}}$.

\section{Decisional Gradient Descent: divergence}
\label{app:DGD_div}
We now show how to obtain the final expression for the divergence used in the Decisional Gradient Descent optimiser from its definition
\begin{align*}
D_{\text{VMC}}(p_\theta, p_{\theta'})
    &=
    \mathbb{E}_{X \sim p_{\theta}}
    \left[  E_{L,\theta}(X) - E_{L,\theta'}(X)\right] \\
    &= \frac{1}{2}\sum_{k=1}^A \mathbb{E}_{X \sim p_\theta} \left [ \partial_{x_k}^2\ln\vert\Psi_{\theta'}\vert +\left( \partial_{x_k}\ln\vert\Psi_{\theta'}\vert\right)^2 - \partial_{x_k}^2\ln\vert\Psi_{\theta}\vert - \left( \partial_{x_k}\ln\vert\Psi_{\theta}\vert\right)^2\right ] \, .
\end{align*}
Now, using the fact that $p_\theta=\vert\Psi_\theta\vert^2$, we have
\begin{align*}
    D_{\text{VMC}}(p_\theta, p_{\theta'}) &= \frac{1}{8}\sum_{k=1}^A \mathbb{E}_{X \sim p_\theta} \left[ 2\partial_{x_k}^2\left( \ln{p_{\theta'}-\ln{p_\theta}}\right) + \left( \partial_{x_k}\ln{p_{\theta'}} \right)^2 - \left( \partial_{x_k}\ln{p_{\theta}} \right)^2 \right] \\
    &= \frac{1}{8}\sum_{k=1}^A \mathbb{E}_{X \sim p_\theta} \left\{ 2 \left [ \frac{\partial_{x_k}^2 p_{\theta'}}{p_{\theta'}} - \left( \partial_{x_k}\ln{p_{\theta'}} \right)^2 - \frac{\partial_{x_k}^2 p_{\theta}}{p_{\theta}} + \left( \partial_{x_k}\ln{p_{\theta}} \right)^2 \right] \right. \\
    & + \left. \left( \partial_{x_k}\ln{p_{\theta'}} \right)^2 - \left( \partial_{x_k}\ln{p_{\theta}} \right)^2 \right\} \\
    &= \frac{1}{8}\sum_{k=1}^A \mathbb{E}_{X \sim p_\theta} \left[ 2\frac{\partial_{x_k}^2 p_{\theta'}}{p_{\theta'}} - 2\frac{\partial_{x_k}^2 p_{\theta}}{p_{\theta}} - \left( \partial_{x_k}\ln{p_{\theta'}} \right)^2 + \left( \partial_{x_k}\ln{p_{\theta}} \right)^2 \right] \, .
\end{align*}
We shall now compute the first two terms inside the square bracket individually for a given $k$. For the first one can integrate by parts so that
\begin{align*}
    \mathbb{E}_{X \sim p_\theta} \left[ \frac{\partial_{x_k}^2 p_{\theta'}}{p_{\theta'}} \right] &= \int d(X\backslash \{x_k\})\int dx_k \frac{p_\theta (X)}{p_{\theta'}(X)}\partial_{x_k}^2 p_{\theta'}(X) \\
    &=  \left[ \frac{p_\theta (X)}{p_{\theta'}(X)} \partial_{x_k} p_{\theta'}(X) \right]_{-\infty}^{+\infty} - \int dX \partial_{x_k}p_{\theta'}(X)\frac{p_{\theta'}\partial_{x_k}p_\theta - p_\theta \partial_{x_k}p_{\theta'}}{p_{\theta'}^2} \\
    % &=  \lim_{x_k\to\infty} \left[ \frac{p_\theta (X)}{p_{\theta'}(X)} \partial_{x_k} p_{\theta'}(X) \right]_{-x_k}^{x_k} - \int dX \partial_{x_k}p_{\theta'}(X)\frac{p_{\theta'}\partial_{x_k}p_\theta - p_\theta \partial_{x_k}p_{\theta'}}{p_{\theta'}^2} \\
    &= - \mathbb{E}_{X\sim p_\theta}\left[ \left( \partial_{x_k}\ln{p_{\theta'}}\right)\left( \left( \partial_{x_k}\ln{p_{\theta}}\right) - \left( \partial_{x_k}\ln{p_{\theta'}}\right) \right) \right] \\
    &= \mathbb{E}_{X\sim p_\theta} \left[ \left( \partial_{x_k}\ln{p_{\theta'}} \right)^2 - \left( \partial_{x_k}\ln{p_{\theta'}}\right)\left( \partial_{x_k}\ln{p_{\theta}}\right)\right]\, .
\end{align*}
Notice that in the second line we use the fact that the boundary term arising from the integration by parts is null. While we have checked analytically that this is the case for our neural network ansatz, there is no reason to assume that this will also be the case for arbitrary ans\"atze. However, even if it was non zero, one could compute it without increasing the computational complexity of the overall algorithm.
Regarding the second therm, we have
\begin{align*}
    \mathbb{E}_{X \sim p_\theta} \left[ \frac{\partial_{x_k}^2 p_{\theta}}{p_{\theta}} \right] &= \int d(X\backslash \{x_k\})\int dx_k \partial_{x_k}^2 p_{\theta}(X) \\
    &= \int d(X\backslash \{x_k\}) \left[ \partial_{x_k}p_\theta(X) \right]^{+\infty}_{-\infty} = 0 \, ,
    % \lim_{x_k\to\infty}\partial_{x_k}p_\theta(X) - \lim_{x_k\to -\infty}\partial_{x_k}p_\theta(X)  \right) = 0 \, .
\end{align*}
where we have used the fact that the derivatives of the probability distribution of our NQS vanish at infinity.
Finally, plugging these two terms into the expression of the divergence we obtain the desired expression for the divergence, namely
\begin{align}
    D_{\text{VMC}}(p_\theta, p_{\theta'}) &= \frac{1}{8}\sum_{k=1}^A \mathbb{E}_{X \sim p_\theta} \left[ \left( \partial_{x_k}\ln{p_{\theta'}} \right)^2 - 2\left( \partial_{x_k}\ln{p_{\theta'}} \right) \left( \partial_{x_k}\ln{p_{\theta}} \right) + \left( \partial_{x_k}\ln{p_{\theta}} \right)^2 \right] \nonumber \\
    &= \frac{1}{8}\sum_{k=1}^A \mathbb{E}_{X \sim p_\theta} \left[ \left( \partial_{x_k}\ln{p_\theta} - \partial_{x_k}\ln{p_{\theta'}} \right)^2 \right] \, . \label{eq:VMCdiv}
\end{align}

\section{Decisional Gradient Descent: metric}
\label{app:DGD_metric}
As stated in the main text, the metric associated to a given divergence can be obtained by performing a Taylor expansion on the divergence around one of the two parametrisations.
We start from Eq.~\eqref{eq:VMCdiv} and expand in $\delta$ as $(\theta, \theta+\delta)$ up to second order.
For this we first Taylor expand $\ln{p_{\theta + \delta}}$ as
\begin{equation*}
    \ln{p_{\theta + \delta}}
        = \ln{p_{\theta}}
            + \sum_{j} \delta_j  \ \partial_{\theta_j} \ln{p_{\theta}}
            + \mathcal{O}(\delta^2) \ ,
\end{equation*}
so that
\begin{equation*}
    \left( \partial_{x_k}\ln{p_\theta} - \partial_{x_k}\ln{p_{\theta'}} \right)^2
        = \sum_{ij} \delta_i  \ \delta_j \times (\partial_{\theta_i} \partial_{x_k} \ln{p_{\theta}}) \ (\partial_{\theta_j} \partial_{x_k} \ln{p_{\theta}})
            + \mathcal{O}(\delta^2) \ .
\end{equation*}
Then we recover the Taylor expansion of the divergence as
\begin{align*}
    D_{\text{VMC}}(p_\theta, p_{\theta + \delta}) &= \frac{1}{8}\sum_{k=1}^A \sum_{ij} \delta_i  \ \delta_j \times 
    \mathbb{E}_{X \sim p_{\theta}} 
\left[ 
    (\partial_{\theta_i} \partial_{x_k} \ln{p_{\theta}}) \ (\partial_{\theta_j} \partial_{x_k} \ln{p_{\theta}})
\right ] + \mathcal{O}(\delta^2) \\
    &= \frac{1}{2}\delta^\transpose
    \underbrace{\left[\frac{1}{4} \sum_{k=1}^A \mathbb{E}_{X \sim p_{\theta}} \left[ (\nabla_{\theta} \partial_{x_k} \ln{p_{\theta}}) \ (\nabla_{\theta} \partial_{x_k} \ln{p_{\theta}})^\transpose\right]\right]}_{G_\text{VMC}(\theta)} \delta
    + \mathcal{O}(\delta^2) \, ,
\end{align*}
and identify the metric $G_{\text{VMC}}(\theta)$ with the equation given in the main text, i.e.\
\begin{equation*}
    G_{\text{VMC}}(\theta)_{\theta_i \theta_j}
    \equiv
    \frac{1}{4}
    \sum_{k=1}^A
    \mathbb{E}_{X \sim p_\theta}
    \left[
        \partial_{\theta_i}\partial_{x_k} \ln p_{\theta}(X) \ 
        \partial_{\theta_j}\partial_{x_k} \ln p_{\theta}(X)
    \right] \, .
\end{equation*}

\section{Performance of Adam vs DGD}

\begin{figure*}
\includegraphics[width=0.94\linewidth, trim=180 80 180 0]{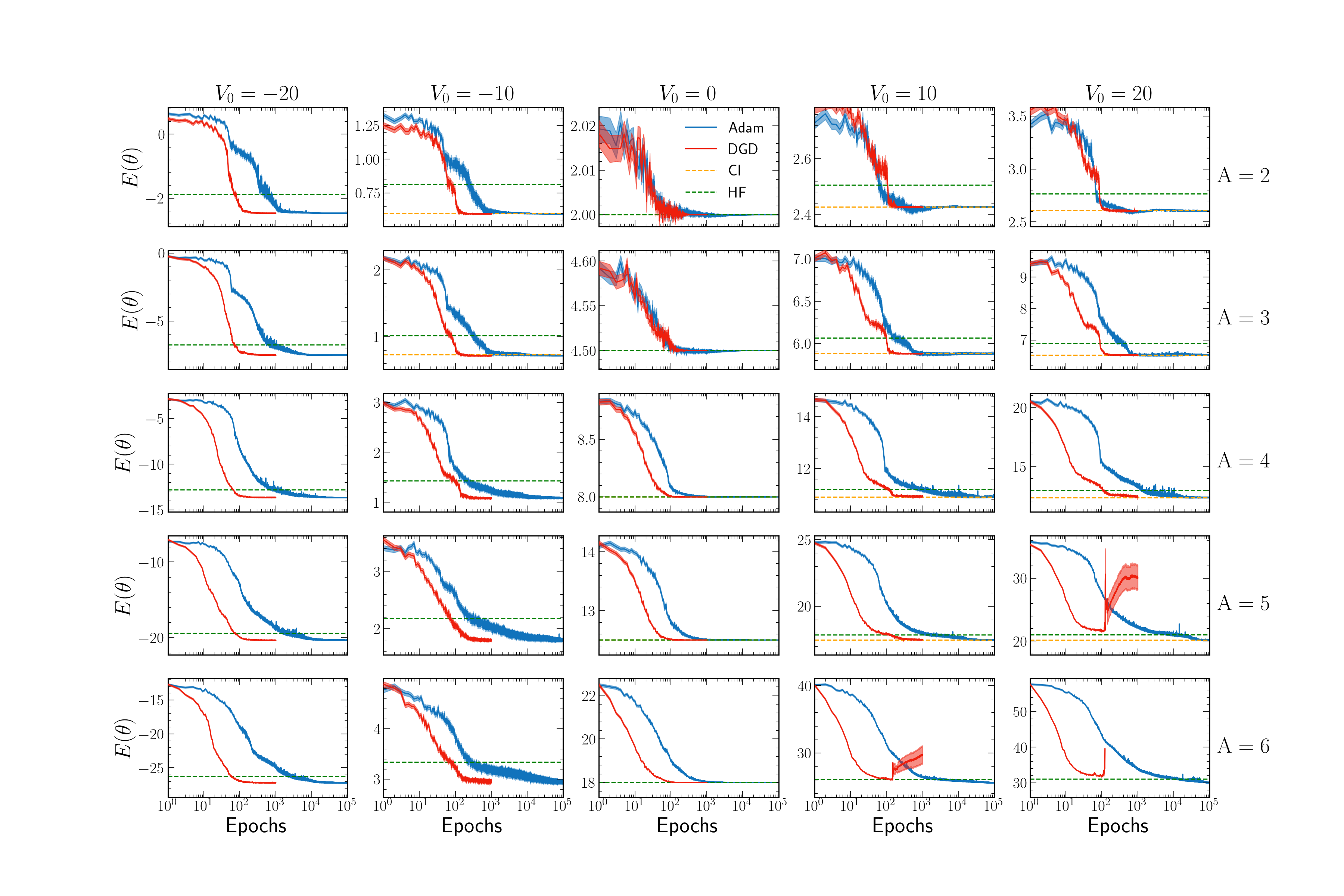}
	\caption{
 The same as Fig.~3 in the main text, but comparing the DGD optimiser (red solid line)
 to Adam (blue solid line). 
 Note the logarithmic scale in the $x-$axis.}
	\label{fig:DGD_vs_Adam}
\end{figure*}

In the main text, we comment briefly on the performance of DGD compared to the standard Adam optimizer.
We provide here a more systematic analysis.
We plot the energy as a function of the epoch number 
 for $A \in \set{2,3,4,5,6}$ and $V_0 \in \set{-20,-10,0,10,20}$ in 
 Fig.~\ref{fig:DGD_vs_Adam}.
 We use a log scale for the $x$-axis for clarity.
 Overall, we observe a substantial decrease in the number 
 of epochs required
 to reach convergence.
 Depending on the system, this decrease is by a factor of
 $10$ or up to a $100$.
 We also confirm that such acceleration is emphasized 
in the late stage of convergence. 
This is a typical feature of second-order 
 optimisers~\cite{martens2012training,martens2020new}.

Our simulations show two additional relevant features. 
First, concerning Adam, we find that the number of epochs
required to reach a converged result tends to increase
with particle number (e.g. as one moves down in the rows
of Fig.~\ref{fig:DGD_vs_Adam}). 
For instance, Adam converges in about $10^3$ epochs
for $A=2$, whereas it requires $\sim 10^5$ epochs in the
$A=6$ systems.
Second, and in stark contrast to the previous observation, 
the performance of DGD stays relatively constant as $A$ increases.
All simulations are converged within about $10^2$ epochs. 
We take this improvement of performance as a sign of the adaptability of our optimiser.
This is another typical empirical feature of second-order optimisers.

\section{Neural Network and Hyperparameters}

We provide here detailed information regarding the architecture
and the hyperparameters of our NQS simulations. 
The architecture of our network is discussed in detail in Refs.~\cite{keeble2022neural,keeble2023machine}.
The network architecture parameters and hyperparameters are listed in Table~\ref{tab:network}.

\begin{table}[H]
  \begin{center}
    \begin{tabular}{l||l} 
      \hline
      \textbf{Network Hyperparameters} & \textbf{Value}\\
      \hline
      Number of hidden nodes, $H$ & $64$ \\ 
      Number of equivariant layers, $L$ & $2$ \\
      Number of determinants & $1$ \\
      Activation function & $\tanh(x)$ \\
      Weight initialisers &  Kaiming uniform \cite{he2015delving} \\
      Bias initialisers & Kaiming uniform \cite{he2015delving} \\
      Envelope structure & Origin-centered Gaussian \\
      \hline
    \end{tabular}
    \caption{List of network hyperparameters.}
    \label{tab:network}
  \end{center}
\end{table}

We further list the settings of our MCMC sampling in Table~\ref{tab:MCMC}.

\begin{table}[H]
  \begin{center}
    \begin{tabular}{l||l} 
      \hline
      \textbf{Sampling Hyperparameters} & \textbf{Value}\\
      \hline
      Number of walkers, $N_W$ & $4096$ \\ 
      Initial distribution of walkers & $\mathcal{N}(0,1)$ \\
      MCMC steps per epoch & $400$ \\
      Local energy clipping constant (Ref.~\cite{pfau2020ab}) & $5$ \\
      Target Metropolis-Hastings acceptance rate & $50\%$ \\
      \hline
    \end{tabular}
    \caption{List of sampling hyperparameters.}
    \label{tab:MCMC}
  \end{center}
\end{table}

For each simulation, we first pre-train the NQS by minimising its overlap with a set of target orbitals.
The target orbitals are chosen to be Hermite polynomials, which are the exact solutions for the non-interacting Harmonic Oscillator case~\cite{keeble2023machine}.
The initial NQS and Monte Carlo walkers are identical between different optimisers in order to avoid initialisation effects in the convergence.
We further list the settings of our pre-training optimisation in Table~\ref{tab:pretraining}.

\begin{table}[H]
  \begin{center}
    \begin{tabular}{l||l} 
      \hline
      \textbf{Pre-training Hyperparameters} & \textbf{Value}\\
      \hline
      Pre-training optimiser & Adam \\
      Learning rate, $\alpha$ & $10^{-4}$ \\
      First-order exp. decay rate, $\beta_1$ & $0.9$ \\
      Second-order exp. decay rate, $\beta_2$ & $0.999$ \\
      Divide-by-zero constant, $\varepsilon$ & $10^{-8}$ \\
      Pre-training loss function & Mean Squared Error (Eq.~(20) of Ref.~\cite{keeble2023machine}) \\
      Pre-training orbitals & Hermite polynomials \\
      Number of epochs & $10^3$ \\
      \hline
    \end{tabular}
    \caption{List of pre-training hyperparameters.}
    \label{tab:pretraining}
  \end{center}
\end{table}

All our optimisers share a set of common hyperparameters, shown at the top of Table~\ref{tab:optimisers}.
The additional hyperparameters for each individual optimiser are shown in the subsequent rows of Table~\ref{tab:optimisers}.
In the cases where MINRES is employed, we use the MINRES implementation imported from the SciPy package~\cite{SciPy,paige1975solution}.

\begin{table}
  \begin{center}
    \begin{tabular}{l||l} 
      \hline
      \textbf{Common Hyperparameters} & \textbf{Value}\\
      \hline
      Damping adaptive interval, $T$ & $5$ \\
      Damping (scaling) regularisation scheme & Diagonal shift (Eq.~3.24a) \\
      Damping (scaling) adaptive scheme & Levenberg-Marquardt rule (Eq.~3.25) \\
      Initial damping (scaling), $\lambda_0$ & $10^3$ \\
      Damping (scaling) adaptive constant, $\omega_1$ & $(19/20)^{5}$ \\
      Minimum damping (scaling), $\lambda_{\text{min}}$ & $10^{-7}$ \\
      Maximum damping (scaling), $\lambda_{\text{max}}$ & $10^{20}$ \\
      % L2 regularisation, $\eta$ & 0 \\
      \hline
      \textbf{KFAC Hyperparameters} & \textbf{Value}\\
      \hline
      Damping (direction) regularisation scheme & Factored Tikhonov  (Sec.~E.6 of Ref.~\cite{martens2015optimizing}) \\
      Damping (direction) adaptive scheme & Greedy algorithm (Sec.~E.6 of Ref.~\cite{martens2015optimizing}) \\
      Initial damping (direction), $\gamma_0$ & $\sqrt{10^3}$ \\
      Damping (direction) adaptive constant, $\omega_2$ & $\sqrt{19/20}^{5}$ \\
      Minimum damping (direction), $\gamma_{\text{min}}$ & $10^{-4}$ \\
      Maximum damping (direction), $\gamma_{\text{max}}$ & $10^{20}$ \\
      Maximum exp. moving average constant, $\bar{\epsilon}$ & $0.9$ (Sec.~D of Ref.~\cite{martens2015optimizing})\\ 
      Preconditioning matrix (direction) & $\breve{F}_{\text{KFAC}}$ (Eq.~3.21) \\
      Quadratic Model (rescaling) & $F$ (Eq.~3.11) \\
      \hline
      \textbf{QN-(MR)-KFAC Hyperparameters} & \textbf{Value} \\
      \hline
      Damping (direction) regularisation scheme & Factored Tikhonov \\
      Damping (direction) adaptive scheme & Greedy algorithm \\
      Initial damping (direction), $\gamma_0$ & $\sqrt{10^3}$ \\
      Damping (direction) adaptive constant, $\omega_2$ & $\sqrt{19/20}^{5}$ \\
      Minimum damping (direction), $\gamma_{\text{min}}$ & $10^{-4}$ \\
      Maximum damping (direction), $\gamma_{\text{max}}$ & $10^{20}$ \\
      Number of MINRES iterations, $N_\text{MR}$ & $0$, $50$ (QN-KFAC/QN-MR-KFAC)\\
      MINRES starting point & $\Delta^{\text{KFAC}}_n$ (Eq.~3.22)\\
      MINRES matrix & $\breve{F}$ (Eq.~3.21) \\
      MINRES preconditioner parameters & $\xi = 0.75$ and $\kappa = 10^{-2}$ \\
      MINRES preconditioner matrix & $(\text{diag}(\breve{F}) + \kappa I )^\xi$\\
      Quadratic Model (rescaling) & $H_Q$ (Eq.~3.29) \\
      \hline
      \textbf{NGD / DGD Hyperparameters} & \textbf{Value} \\
      \hline
      Damping (direction) regularisation scheme & Diagonal shift\\
      Damping (direction) adaptive scheme & Levenberg-Marquardt rule\\
      Initial damping (direction), $\lambda_0$ & $10^3$ \\
      Damping (direction) adaptive constant, $\omega_1$ & $(19/20)^{5}$ \\
      Minimum damping (direction), $\lambda_{\text{min}}$ & $10^{-7}$ \\
      Maximum damping (direction), $\lambda_{\text{max}}$ & $10^{20}$ \\
      Number of MINRES iterations, $N_\text{MR}$ & $50$ \\
      MINRES starting point & $\zeta \times \delta_{n-1}$, with $\zeta = 0.95$\\
      MINRES matrix & $\breve{F}$, $\breve{G}_{\text{VMC}}$ (NGD/DGD)\\
      MINRES preconditioner parameters & $\xi = 0.75$ and $\kappa = 10^{-2}$ \\
      MINRES preconditioner matrix & $(\text{diag}(\breve{F}) + \kappa I )^\xi$, $(\text{diag}(\breve{G}_{\text{VMC}}) + \kappa I )^\xi$ (NGD/DGD)\\
      Quadratic Model (rescaling) & $F$, $G_{\text{VMC}}$ (NGD/DGD)\\
      \hline
    \end{tabular}
    \caption{List of hyperparameters for all optimiser of the main paper with the common hyperparameters stated at the top of the table.
    All additional hyperparameters for each optimiser are stated within their own subsection in the table.}
    \label{tab:optimisers}
  \end{center}
\end{table}

\hphantom{a} \newpage

\bibliographystyle{RS}
\bibliography{bibliography}

%%% Make sure to upload the bib file along with the tex file and PDF
%%% Please see the test.bib file for some examples of references